\documentclass{aa}

\usepackage{graphicx}
\usepackage{natbib}
\usepackage{longtable}
\usepackage{booktabs}
\usepackage{rotating}
\usepackage{txfonts}
\usepackage{textcomp}
\begin{document}

\title{A survey for low-mass stellar and substellar members of the Hyades open cluster.}

\author{Stanislav Melnikov \inst{1,2} \and Jochen Eisl\"offel \inst{1}}

\institute{Th\"uringer Landessternwarte Tautenburg, Sternwarte 5, 07778 Tautenburg, Germany  \and 
           Ulugh Beg Astronomical Institute, Astronomical str. 33, 700052 Tashkent, Uzbekistan}

\offprints{Stanislav Melnikov,
\email{smeln2005@gmail.com}}

\date{Received / Accepted}

\abstract{Unlike young open clusters (with ages $<$ 250 Myr), the Hyades cluster (age $\sim$ 600 Myr) has a clear deficit of very low-mass stars (VLM) 
and brown dwarfs (BD). Since this open cluster has a low stellar density and covers several tens of square degrees on the sky, extended surveys are 
required to improve the statistics of the VLM/BD objects in the cluster.
}
{We search for new VLM stars and BD candidates in the Hyades cluster to improve the present-day cluster mass function down to substellar 
masses.
}
{An imaging survey of the Hyades with a completeness limit of $21\fm5$ in the $R$ band and $20\fm5$ in the $I$ band was carried out with the
2k\,$\times$\,2k CCD Schmidt camera at the 2m Alfred Jensch Telescope in Tautenburg. We performed a photometric selection of the cluster member candidates by
combining results of our survey with 2MASS $JHK_s$ photometry. 
}
{We present a photometric and proper motion survey covering 23.4 deg$^2$ in the Hyades cluster core region. Using optical/IR colour-magnitude diagrams, we 
identify 66 photometric cluster member candidates in the magnitude range $14\fm7<I<20\fm5$. The proper motion measurements are based on several all-sky 
surveys with an epoch difference of 60-70 years for the bright objects. The proper motions allowed us to discriminate the cluster members from field objects 
and resulted in 14 proper motion members of the Hyades. We rediscover Hy 6 as a proper motion member and classify it as a substellar 
object candidate (BD) based on the comparison of the observed colour-magnitude diagram with theoretical model isochrones.
}
{With our results, the mass function of the Hyades continues to be shallow below $\sim0.15 M_\odot $ indicating that the Hyades have 
probably lost their lowest mass members by means of dynamical evolution. We conclude that the Hyades core represents the `VLM/BD desert' and that most 
of the substeller objects may have already left the volume of the cluster.
}

\titlerunning{The VLM/BDs in the Hyades cluster}
\maketitle

\section{Introduction}
\label{intro} 
The accurate initial mass function (MF) of Galactic open clusters allows us to build up a picture of the initial conditions of cluster formation and to 
investigate their further evolution. The bright end of the mass function has been analysed in many detailed studies of bright clusters. In the last decades, on 
the other hand, deep photometric surveys of open clusters were focused on the faint MF end reaching out to the lowest mass members. The nearby open clusters are 
very convenient targets for this goal. In the solar  neighbourhood there are a number of open clusters including the Pleiades, the Hyades, Praesepe (M44), and 
the Coma Berenices open cluster (Melotte 111). Extensive deep surveys of the Pleiades (age $\sim$120 Myr) have led to the discovery of a large 
population of very low-mass stars (VLM) and substellar members known as brown dwarfs (BDs) \citep{Sta07,Lod12a,Zap14,Bouy15}. A considerable low-mass 
population has also been discovered in some other younger stellar clusters (\citealp[$\alpha$ Per:][]{Lod12b}; \citealp[$\sigma$ Ori:][]{Pena12}). 
These surveys showed that the initial conditions of star formation in stellar clusters were effective in creating low-mass members.

However, the proximity of a cluster also has the disadvantage of a large extension on the sky, which renders surveys for cluster membership difficult. 
Moreover, with cluster evolution the lowest mass members may escape from a cluster core (`evaporate' from the cluster) due to dynamical encounters 
and mix-up with the field objects \citep{Terl87,Kr95,Fue00}. Therefore, detection of these objects will be more  complicated. \Citet{Fue00} 
suggested that this effect can already be noticeable in the clusters with ages of $\geq$200 Myr. The first surveys of the intermediate-age clusters, 
which covered only a small percent of the cluster core areas, did  not find any significant population of low-mass members, similar to what was found 
for the Pleiades. The first studies of the Coma open cluster (age$\sim500$ Myr) showed that this cluster has a deficit of low-mass stars in comparison 
to the younger clusters \citep{Del81,Oden98}. The recent and deeper surveys of this cluster covered a larger field and were reaching into the 
substellar domain. They confirmed that this deficit seems to be intrinsic and this finding \citep{Cas06,Mel12} supports the idea that the depletion is 
caused by dynamical evolution. Recent studies of Praesepe ($\sim$600 Myr), another open cluster of similar age, based on the analysis of all-sky 
surveys UKIDSS \citep{Boud12} and 2MASS, PPMXL, Pan-STARRS \citep{Wang14} found conflicting results. \citet{Boud12} found that the Praesepe MF is 
consistent with that of the Galactic disk population down to $0.07\,M_\odot$, whereas \citet{Wang14} concluded that the cluster MF shows a deficit of 
members below $0.3\,M_\odot$. Therefore, deep wide-field surveys of intermediate-aged open clusters (with ages of 450--600 Myr) are required to 
improve the comparison of their mass functions with those of the younger clusters and construct a more reliable scenario of how open clusters evolve 
with age.

The \object{Hyades} open cluster (d=46 pc) with age $\sim$600 Myr \citep[age=625 Myr,][]{Per98}, which has a Pleiades-like stellar population, is one 
of the most studied open clusters located in the solar neighbourhood. The earliest wide-field search for Hyades members covering $\sim$110 deg$^2$ 
discovered a deficit of M-type dwarfs in this cluster \citep{Reid92,Reid93} with respect to the solar neighbourhood mix. From these observations 
\citet{Reid92} derived a gravitational binding radius of $\sim$10.5 pc and a total cluster mass of 410--480\,$M_\odot$. Deeper imaging surveys, which 
covered only small areas to reach fainter members did not lead to a discovery of any significant population of such objects 
\citep{Reid99,Reid00,Giz99}. A survey covering 10.5 deg$^2$ by \citet{Dob02} also failed to find any new low-mass members and recovered only the known 
stellar member RHy 29 \citep{Reid93}. A wide-field study of the Hyades based on the recent proper motion catalogue PPMXL, 2MASS, and  Carlsberg 
Meridian Catalogue 14  \citep[CMC14;][]{Cop06} photometry has enabled a full census of the kinematic cluster members down to masses of 
$\sim$0.2\,$M_\odot$ in a region up to 30 pc from the cluster  centre \citep{Ros11}. Combining these three surveys, \citet{Ros11} carried out a 
three-dimensional analysis of the cluster population and found that the Hyades have a tidal radius of $\sim$9 pc with clear outward mass segregation. 
Using the PPMXL and Pan-STARRS1 sky surveys \citet{Gold13} analysed the same area of the Hyades and pushed the census of its members down to 
$0.09\,M_\odot$. Based on the deepest survey over 16 deg$^2$ of the Hyades core, \citet{Bou08} reported the discovery of the first two BDs and 
confirmed membership of 19 low-mass stellar members. Analysing the statistically significant number of Hyades members found in the previous studies, 
\citet{Bou08} concluded that the present-day mass function of the Hyades is clearly deficient in the VLM/BD domain compared to the initial MF of the 
Pleiades, which have a similar population structure, but are much younger than the Hyades. Another study based on the all-sky surveys UKIDSS and 2MASS 
\citep{Hog08} and following spectroscopic analysis \citep{Cas14,Lod14} added several BDs to the substellar domain of the cluster. Nevertheless, the 
updated MF still shows the apparent deficit of the lowest mass members \citep{Lod14}.

In this paper, we present the results of a new deep imaging survey of the Hyades open cluster obtained with the wide-field Schmidt camera at the 2m Alfred 
Jensch Telescope of the Th\"uringer Landessternwarte in Tautenburg, Germany (TLS). Details of the photometric observations and data reduction are described in 
Sect. 2. In Sect. 3 we introduce the photometric selection procedure of VLM objects and BD candidates; we then compute proper motions of 
our optically selected candidates by combining the TLS astrometric calibration with earlier epoch all-sky surveys, and describe the results. 
In Sect. 4 we report on the comparison of our photometric selection with the results of previous surveys of the Hyades; we also discuss  the updated mass 
function of the cluster combining our new cluster member candidates with those from the previous studies.

\section{TLS photometric survey and basic CCD reduction}
\label{obs}
\begin{figure}[t]
\centering
\resizebox{\hsize}{!}{\includegraphics{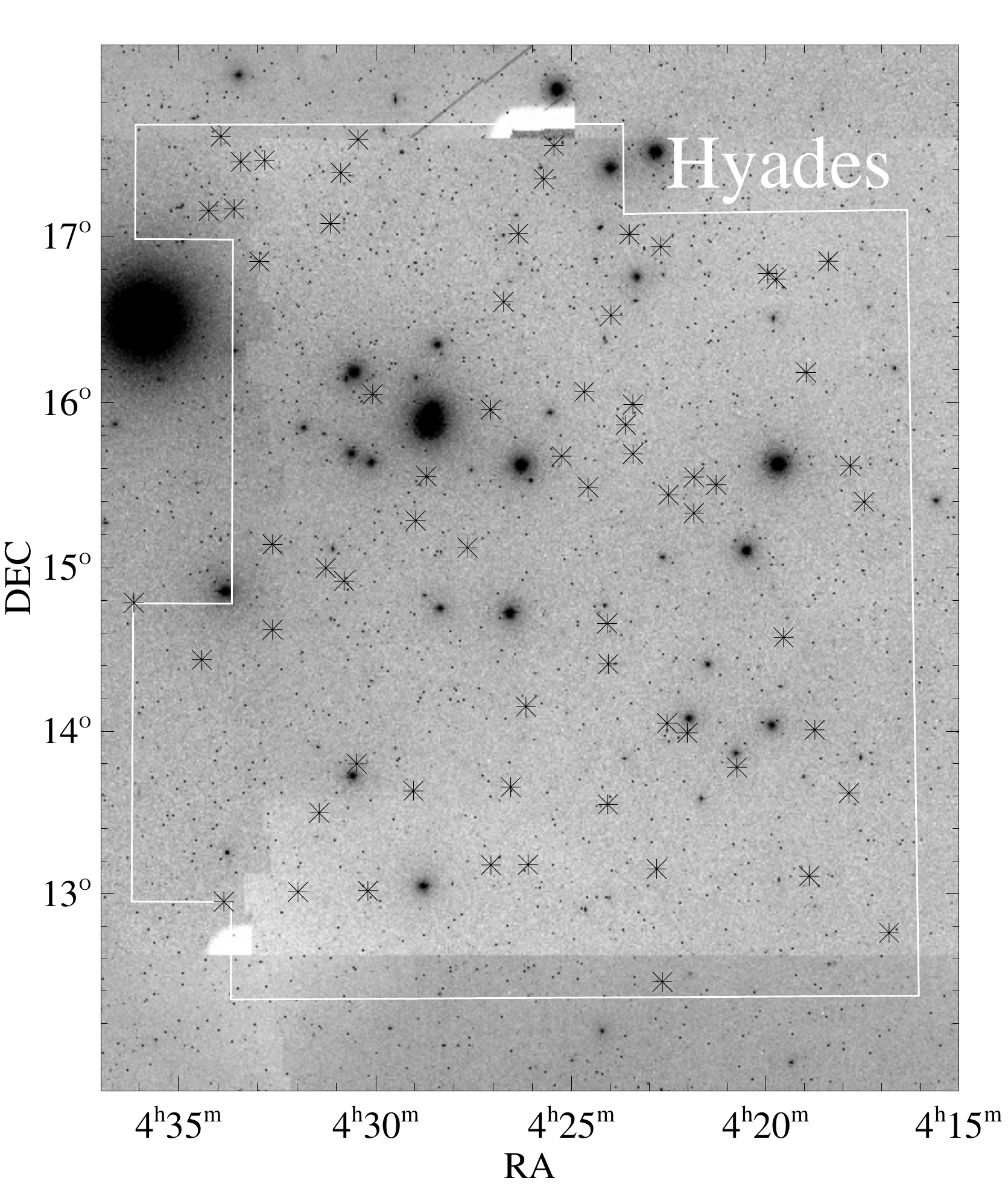}}
\caption{\label{area} Area of the Hyades cluster mapped by our TLS imaging survey. The total size of the surveyed area is $\sim$23.4 deg$^2$. Star symbols are 
photometrically selected Hyades member candidates, which are listed in Table~\ref{phot_cand}.}
\end{figure}

We have carried out a new wide and deep imaging survey of the Hyades open cluster (RA=4$^\mathrm{h}$26$^\mathrm{m}$ DEC=+15\degr\  0\arcmin) in the 
$RI$ bands using the 2k\,x\,2k CCD camera in the Schmidt focus of the 2m Alfred Jensch Telescope in Tautenburg. A short description of the $RI$ 
photometric system of the CCD camera can be found in \citet{Mel12}. The Hyades are a very close open cluster and its core area  \citep[$r\sim2.8$ 
pc;][]{Per98} extends over a very large sky area of $\sim$50 deg$^2$. Our photometric survey obtained in October-November  2006 covers 23.4 deg$^2$ in 
its central area, i.e. about 47\% of the cluster core (see Fig.~\ref{area}).

The total exposure time per filter (one frame) was 600\,s, and for this exposure time the limiting magnitude of the frames was estimated to be $22.5$ 
in the $I$ band. To avoid saturated stellar profiles, we ensured that $I$-magnitudes were in general $>$14.5. In total, 65 fields of good quality were 
obtained in the course of this survey. The positions of the field centres were chosen so that each field overlaps with its neighbours by 
$\sim12^\mathrm{s}$ ($\sim$3\arcmin) in right ascension and $\sim3\farcm5$ in declination. The overlapping regions allowed us to check the quality of 
our photometric calibration by comparing the magnitudes of stars located in the overlapping areas of the adjacent frames.

The raw images were reduced following standard recipes in IRAF\footnote{IRAF is distributed by National Optical Astronomy Observatories, which is 
operated by the Association of Universities for Research in Astronomy, Inc., under contract with the National Science Foundation.}; this procedure 
included overscan correction, bias subtraction, and dome flat-fielding. The $I$-band images contained a prominent interference fringe pattern caused 
by night sky emission.  These fringe strips were removed with a fringe mask constructed from the whole set of the $I$-band images. The $R$- and 
$I$-band images were then aligned where necessary and all were astrometrically calibrated using the Graphical Astronomy and Image Analysis Tool 
software (GAIA) and the Hubble Guide Star Catalog (GSC v.1.2) as  reference. The GSC contains positions for most of the field stars down to magnitude 
$V=16$. Each of our Hyades CCD frames  contains more than 30 reference stars evenly distributed across the field, and the accuracy of the astrometric 
solution for individual images is better than $0\farcs4$ in both filters for both equatorial coordinates.

Cosmic ray hits (cosmics) were removed by combining each pair of $RI$ frames of the same sky field into one image and rejecting cosmics. The images also had 
several different kinds of artefacts  and extended objects which had to be discriminated from star-like objects. A description of these various 
artefacts and 
the method allowing us to clean the images is described in \citet{Mel12}.

Instrumental magnitudes of all extracted sources were then extracted based on the measurement of the point spread function (PSF) using the 
\textit{daophot} package of IRAF. Finally, we converted the instrumental magnitudes into $RI$-band magnitudes using photometric standards observed in 
Landolt Selected Areas \citep{Lan92}. Photometric errors for the $RI$ bands depend on magnitude: in the range of $14-18$ the errors gradually 
increase from $0\fm01$ to $0\fm04$, but for objects with $R>18$ and $I>18$ the errors grow faster and reach about $0\fm3$ for $R\approx20$. For the 
$I-(I-J)$ and $I-(I-K)$ CMDs, the $I$-errors are combined with the infrared photometry errors provided in the 2MASS catalogue.

\section{Selection of very low-mass stellar and brown dwarf candidates}
\label{select}
\begin{figure*}
%\centering
\sidecaption
\begin{minipage}[c]{12cm}
\resizebox{0.5\hsize}{!}{\includegraphics{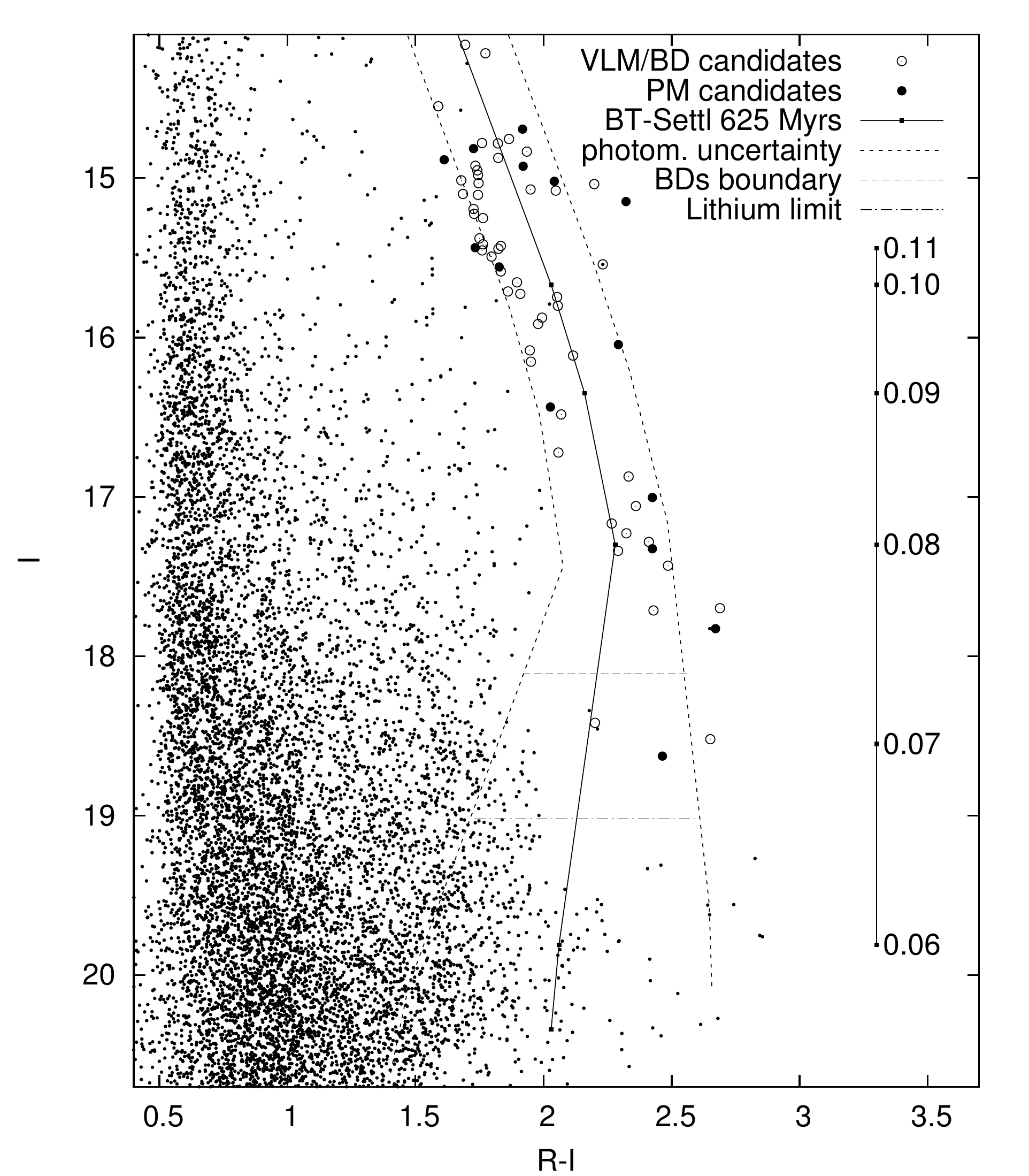}}
\resizebox{0.5\hsize}{!}{\includegraphics{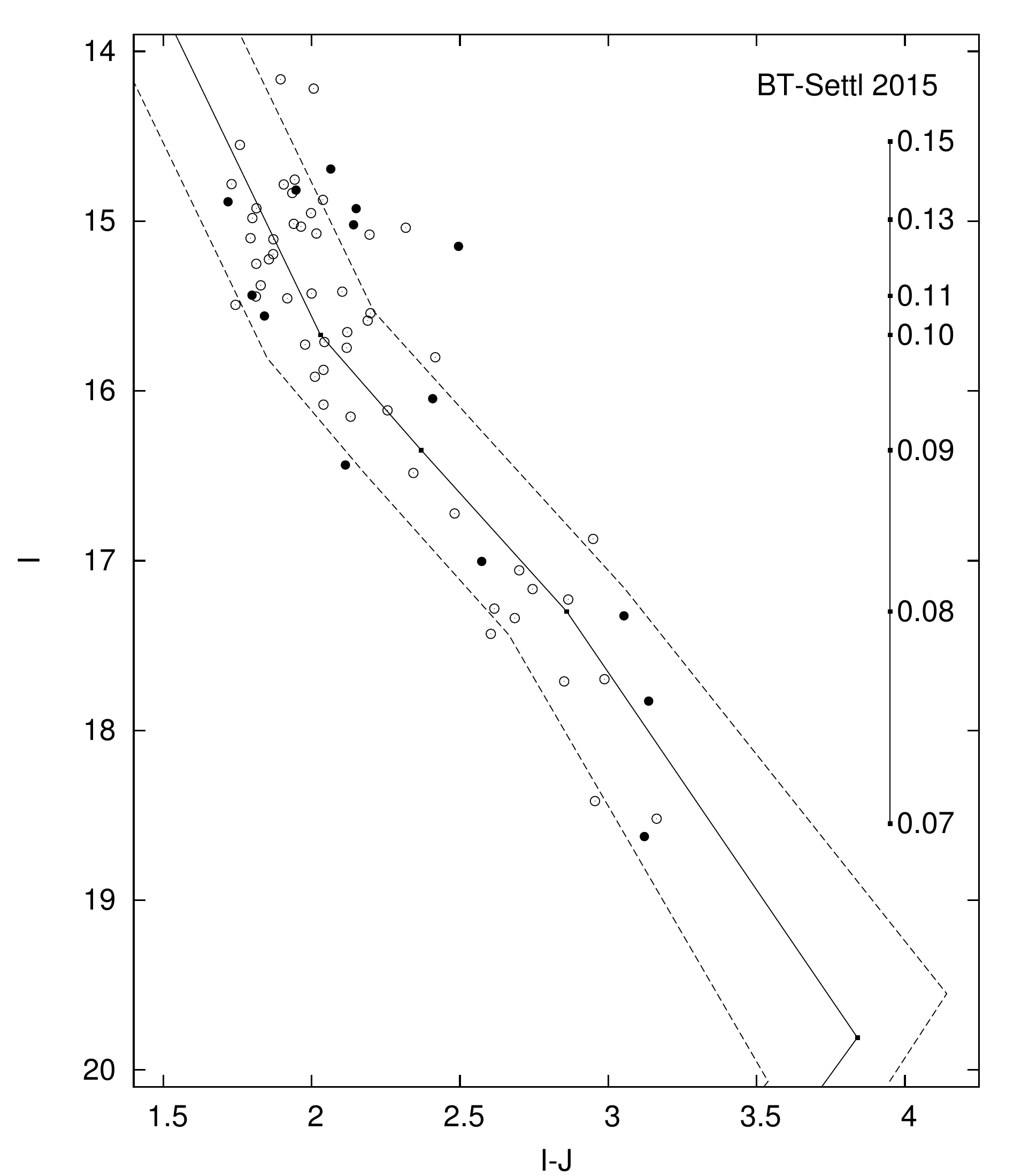}}
\resizebox{0.5\hsize}{!}{\includegraphics{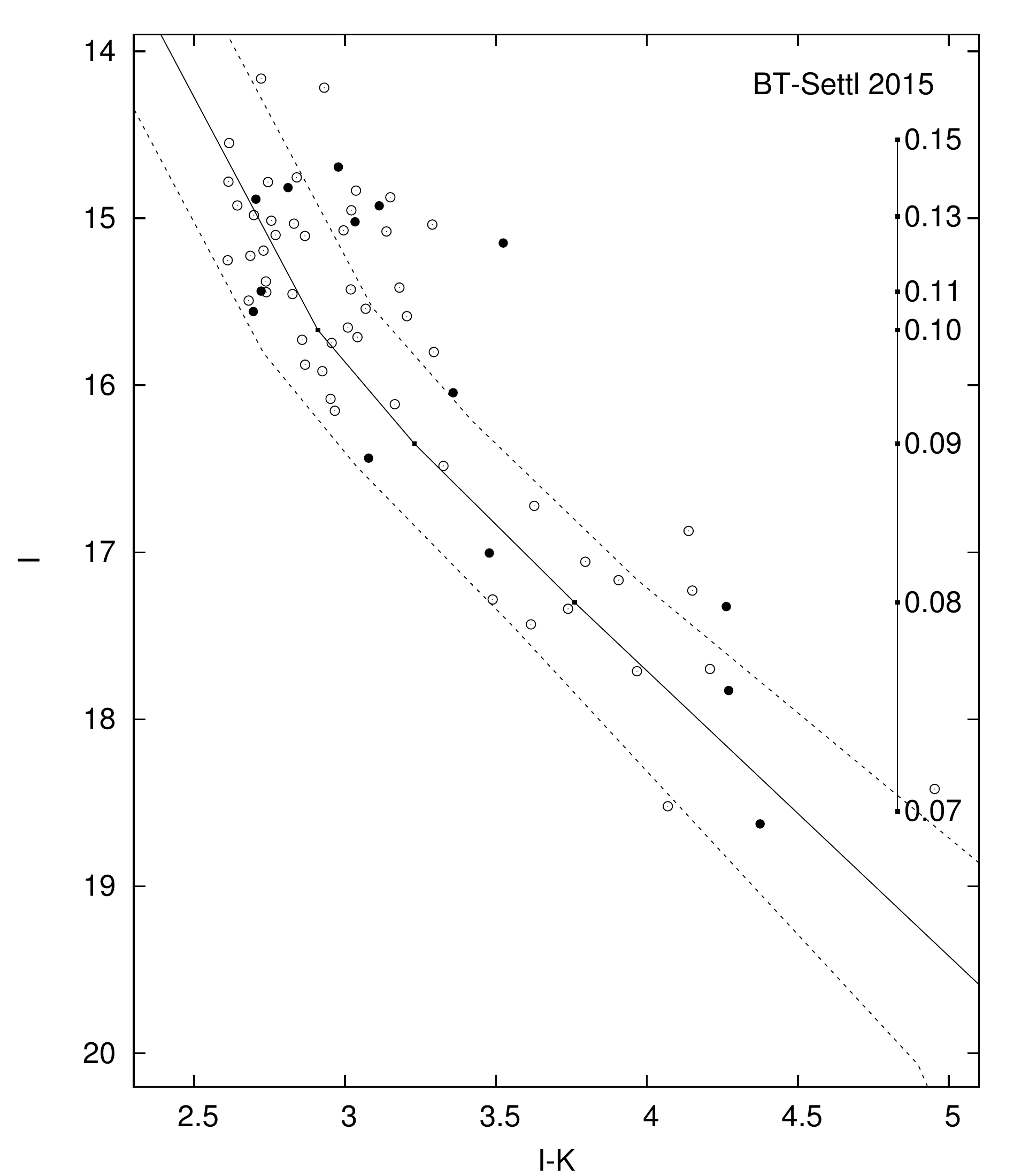}}
\resizebox{0.5\hsize}{!}{\includegraphics{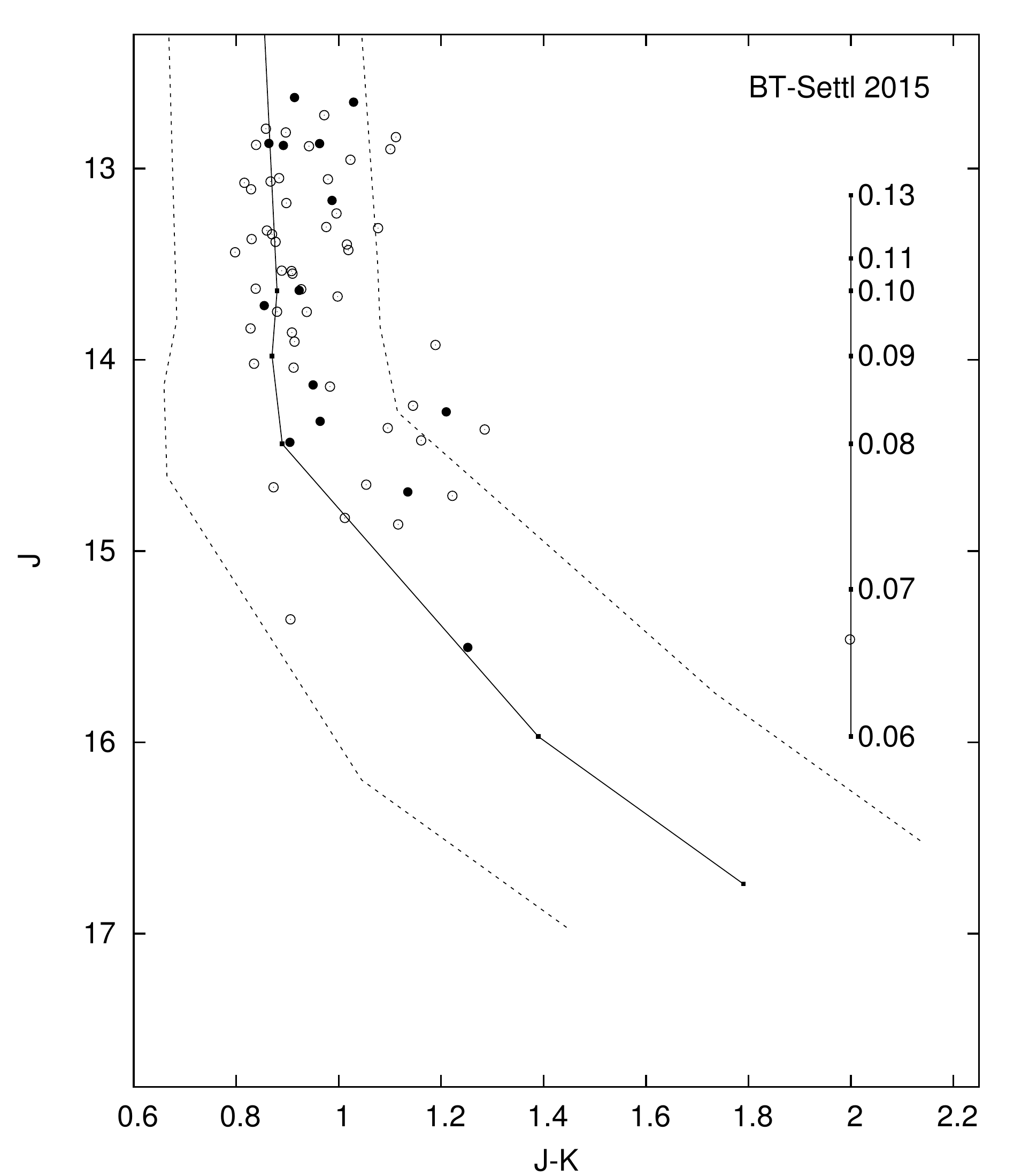}}
\end{minipage}
\caption{\label{CMDs} $I-(R-I)$ (top left), $I-(I-J)$ (top right), $I-(I-K)$ (bottom left), and $J-(J-K)$ (bottom right) colour-magnitude diagrams of 
optically selected candidates (points) from the TLS and 2MASS/UKIDSS surveys; the proper motion candidates are indicated by  filled circles. In all CMDs, we 
show 625 Myr BT-Settl isochrones shifted to the distance of the Hyades cluster ($m-M=3.33$) as solid lines. The vertical line labelled with stellar mass 
($M_\odot$) is the mass scale according to this model. The two dashed lines outline the selection area (see text). The upper horizontal line at $I=18.1$ 
shows the stellar/substellar boundary and the lower line indicates the lithium burning limit. The BT-Settl isochrone in the $J-(J-K)$ CMD agrees well with 
the 
colour indices of the faintest member candidates.} \end{figure*}

All  65 CCD cluster fields together  contain about 290 000 objects that were detected by SExtractor \citep{Ber96} in the $R,I$ bands. We plot 
the $I-(R-I)$ colour-magnitude diagram (CMD)  for the extracted sources  (Fig.~\ref{CMDs}) and compare their diagram position with the model 
isochrones for low-mass objects, shifted to the distance of the Hyades \citep[$m-M=3.33$,][]{Per98}. Interstellar reddening  towards the 
Hyades is very low \citep{Tayl02} and can be neglected. Previous studies of the Hyades members exploited mostly the NextGen \citep{Bar98}, DUSTY 
\citep{Cha00}, and COND \citep{Bar03} isochrones  \citep[e.g.][]{Dob02,Bou08}. NextGen evolutionary models for solar metallicity based on non-grey 
dust-free atmosphere models described various observed properties of M dwarfs down to the bottom of the main sequence (CMDs, 
spectral types, etc.), whereas the DUSTY and COND models try to reproduce the same properties for BDs, taking into account the possible formation 
and opacity of dust grains in the atmosphere of objects with $T_\mathrm{eff}\lesssim2800$ K. The DUSTY and COND models are different in that the 
latter include effects of rapid gravitational settling of the grains in the lower atmospheric layers below the photosphere. \citet{Cha00} predict that 
this process will occur at a temperature of $T_\mathrm{eff}\lesssim1300$\,K, which corresponds to a mass of $M \approx 0.04\,M_\odot$. The BT-Settl 
models \citep{All12} are a further development of evolutionary models which account for the formation of dusty clouds via a parameter-free cloud model 
\citep[based on the cloud microphysics from][]{Ros78}. Compared to DUSTY, the BT-Settl models include, among other microphysical processes, 
gravitational settling of the dust in the cool BD atmospheres. The BT-Settl models, based on a solar abundance from \citet[][CIFIST2011]{Caf11} were 
already employed for the Hyades member selection in \citet{Gold13}, who used the wide global surveys such as Pan-STARRS1 and SDSS combined with 2MASS 
and WISE infrared survey. The BT-Settl model grid allows a good reproduction of near-infrared (NIR) spectral energy distribution of cool VLMs and BDs 
\citep{All14}. The current BT-Settl model grid \citep{Bar15,All16} covers the stellar parameter range for the low-mass objects with $T_{eff}= 
1200-7000$\,K, and therefore the models are valid for both the lowest mass stars and BDs. We decided to utilise the latest BT-Settl model grid for our 
membership analysis.

To distinguish the Hyades member candidates from foreground dwarfs,  we used the BT-Settl isochrones (solid curves in Fig.~\ref{CMDs}) calculated for 
the Hyades age (625 Myr). For the photometric selection of the candidates, we first took into account the cluster depth;  the cluster core has a 
radius of 2.7 parsecs, which means the objects can be 0\fm12 fainter or brighter than the central cluster isochrone. Moreover, the main sequence of the 
Hyades in the $R-(R-I)$ CMD constructed from the previously known members from \citet{Reid93} shows that it is not just a thin line, but has a width 
of $\sim0.2-0.3$ mag, which cannot be explained as resulting from the photometric errors or the cluster depth. \citet{Reid93} analysed this effect on 
$M_V-(V-I)$, but it is also observable in $M_I-(I-K)$. \citet{Reid93} suggested that this effect can be a sequence of natural dispersion of stellar 
parameters, where the high rate of unresolved binaries amongst the cluster members \citep{Gri88,Reid93} can be partly responsible for this scatter. 
Thus, we used the additional colour strip of $0.15$ mag width in order to take into consideration this dispersion and increase the detection 
probability of real members. As a result, we started with all objects within a strip that takes into account the cluster depth, the dispersion of 
colour indices, plus a $1.5\sigma$ wide strip due to photometric uncertainty (PSF photometric errors only). The strips are shown in the $I-(R-I)$ CMD 
in Fig.~\ref{CMDs} as the two dashed curves on both sides of the BT-Settl isochrone which are getting wider due to increasing photometric errors with 
growing magnitude.

When we use the BT-Settl model for analysis of the $I-(R-I)$ CMD, we can roughly split the procedure into two parts. For the bright objects with 
$I<18$, the BT-Settl isochrone is separated from the bulk of the field dwarfs quite well and we have found only several tens of photometric 
candidates within the photometric errors. There are also a number of objects located redward of our red uncertainty boundary which we included in our 
initial sample, especially in the upper part between $I=14-16.5$. Some of these reddish objects at the CMD top could be background dwarfs, distant 
galaxies, or red giants. The observed displacement towards the red could be partly caused by the depth of the cluster and by the binarity of the 
objects. In the most extreme case, a binary consisting of equal mass stars may be lying up to 0.75 mag above the single star sequence. Since we cannot 
identify the origin of this reddening using only the $I-(R-I)$ CMD, all these reddish objects were initially included in our list as potential cluster 
members. As a result, we identified several tens of low-mass cluster member candidates in the magnitude range $14<I<18$, covering from 0.15\,$M_\odot$ 
to about 0.075\,$M_\odot$. These sources were then cross-identified with $JHK$ 2MASS-photometry and were subjected to further checks, which are 
detailed in the following.

For objects with $I>17.5$, the BT-Settl model predicts that the $R-I$ isochrone has a turnover and fainter objects will have bluer colour indices. 
Therefore, for the candidates with masses below $0.075\,M_\odot$, one can see that the model coverage significantly overlaps with the field dwarfs at 
the $I-(R-I)$ CMD (Fig.~\ref{CMDs}). Using  the BT-Settl isochrone has led to a selection of several thousand objects. These sources were then  
cross-identified against sources in 2MASS and the UKIRT InfraRed Deep Sky Survey (Galactic Clusters Survey, UKIDSS). The UKIDSS survey has a 
deeper detection limit, but its photometric selection contained only the $K$ band for the Hyades. We  discuss this photometric selection in the 
following section.

\subsection{Two Micron All-Sky Survey selection}

As the next step of the photometric selection, we used three NIR CMDs of $I-(I-J)$, $I-(I-K)$, and $J-(J-K)$ (see Fig.~\ref{CMDs}). As for the 
$I-(R-I)$ diagram, the 625 Myr BT-Settl isochrones from \citet{All16} are shown on these IR CMDs in Fig.~\ref{CMDs} as solid lines. The two dashed lines 
at both sides of the isochrones indicate the selection area defined by parameters adopted for $I-(R-I)$ CMD such as the photometric uncertainty, the 
cluster depth, and the natural photometric dispersion. All CMDs in  Fig.~\ref{CMDs} also represent  a scale of stellar mass according to the model. 
Using the CMDs, we then selected all the photometric candidates which agree with the theoretical NIR isochrones within defined selection areas. 
Contrary to $I-(R-I)$, the BT-Settl isochrones in the $I-(I-J)$ and $I-(I-K)$ CMDs predict a gradual increase in the colour indices with decreasing  
stellar mass at least down to 0.05\,$M_\odot$. The BT-Settl $J-(J-K)$ isochrone is almost vertical for $J<14.5$, but for fainter $J$ the colour 
index becomes considerably redder. This behaviour agrees well with the colour indices of the faintest member candidates, which show very red colour 
indices on this CMD (Fig.~\ref{CMDs}).

\citet{Gold13} note that the discrimination between the Hyades cluster sequence and the bulk of the background stars is  better when the wavelength 
difference between the bands is greater. Thus, the $I-(I-K)$ CMD is a good one for this preliminary discrimination. First, we cross-identified all our 
initial candidates selected from the $I-(R-I)$ CMD in the Two Micron All-Sky Survey \citep[2MASS,][]{Skr06}. Using a matching radius of 2\farcs5, we 
derived $JHK_s$ photometry for all our candidates up to $I\approx19.3$. We combined the derived NIR photometry with our $I$-magnitudes into the three 
additional CMDs (see Fig.~\ref{CMDs}). For our targets with $I\gtrsim19.3$, the 2MASS survey did not contain IR photometry and thus we 
cross-identified these against sources in the UKIDSS survey. Although the Hyades are covered by this survey, it provides only $K$-magnitudes ($K_1$) 
for this cluster. Finally, we used the transformation equation from \citet{Hew06} to convert UKIDSS $K_1$ magnitudes to 2MASS $K_s$.

The analysis of the large candidate set on the $I-(I-K)$ CMD showed that almost all UKIDSS and 2MASS sources with $I>17.5$ have $I-K$ bluer than  is 
predicted by the BT-Settl $I-(I-K)$ isochrone, and thus they are probably field dwarfs. Therefore, we excluded all the objects from the further 
analysis. In our results, we identified only seven photometric candidates with $I>17.5$. For $I<17.5$ we selected several tens of candidates and 
rejected those candidates which fail our criterion in any of the other NIR CMDs. As a result, all candidates which have only UKIDSS photometry are 
were rejected.

\subsection{Field object contamination and background giant stars}
\label{contam}

An estimate of the number of contaminating field stars can be obtained from the Besan\c{c}on Galaxy model \citep{Rob03}, which gives star counts depending on 
their brightness, colour, and Galactic coordinates. Using this model, \citet{Gold13} have found that the contamination by field stars is negligible ($<10$\%) 
up to 18 pc of the cluster centre. Since our survey lies within this radius, we will not consider contamination as essential for our conclusions.

Finally, we used the $(J-H)-(H-K)$ diagram to weed out possible background giants from our sample. This type of contaminating objects probably experiences 
interstellar extinction and tends to populate the top left side of the diagram above the sequence of the dwarf stars \citep{Gold13}. In \citet{Mel12} we 
tried to find giants as background contaminants projected onto the Coma open cluster with the help of the analysis of narrowband spectral indices 
\citep{Jon73}, which allowed us to distinguish genuine low-mass members from the far red giants with similar colour indices. However, no giants were found in 
the background of the Coma cluster. The Hyades (at $l=180$, $b=-22.3$) are located in the opposite direction to the Galactic centre and quite high above the 
Galactic plane. Therefore, we do not expect to find a large amount of distant background M-type giants in the direction of the Hyades either, given that this 
cluster is located at high Galactic latitude as well.

As a result of this NIR two-colour analysis, we found nine objects that are located in the CMD region with high extinction. To double check, we calculated 
proper motions of these targets (see Sect.~\ref{pm_calc}). Most of these objects have proper motions around zero which seems to favour the idea that they 
might be background objects. Finally, we excluded these objects from our list of photometric VLM candidates.

\subsection{Spectro-photometric classification}

To determine the spectral types of the VLM candidates, we exploited the method of luminosity--spectral type calibration by \citet{Kra07}. This method 
is based on a large sample of stellar spectral energy distributions (SEDs) and allows us to estimate the spectral type (SpT) of a star using only its 
optical/NIR photometry. The results of this classification are presented in Table~\ref{phot_cand}. The second to last column holds the spectral type 
derived from the TLS $I$-band magnitudes and the last column is the average of the three SpT values determined from 2MASS $JHK_s$. Spectral type 
sequences based on  2MASS photometry are  self-consistent and the SpT estimates derived from these bands agree very well: the SpT values are equal to 
or lie within one spectral subclass. At the same time, the spectral type derived from 2MASS is systematically later than those  calculated from the 
TLS $I$ band. Moreover, this difference increases towards later SpT objects. This may imply that our $I$-band calibration has some bias with respect 
to the 2MASS $JHK_s$-system. Considering this result in more detail, one can say that the SpT determined from the two methods agree well (with the 
difference of a SpT subclass) for most objects whose spectral indices were derived from the 2MASS photometry (58 of the 66 objects), whereas for 8 
objects, SpT (2MASS) values are later by two subclasses and larger than those derived from the TLS photometry.

We should note that this luminosity--spectral calibration sequence is based on the stellar SEDs of VLMs only and is valid for stellar objects with 
$K<13.8$ (masses $\gtrsim0.08\,M_\odot$). Therefore, this calibration method may not be reliable for BD candidates and it may not provide the proper 
spectral classification for objects with SpT about and later than L0.

\subsection{Proper motion as a selection criterion}
\label{pm_calc}

The mean proper motion (PM) of the Hyades cluster is $\sim$100 mas\,yr$^{-1}$ \citep{Bou08}, which is quite different from that of Galaxy field stars 
and therefore can be used for the separation of genuine cluster members from background and foreground objects. It was found that the faint Hyades 
members are located within the octant of PM space between PA=90\textdegree\ and PA=135\textdegree\ \citep{Bry94} with a convergent point at 
PA$\sim$115\textdegree.
\begin{figure}
\centering
%\begin{minipage}[c]{14cm}
\resizebox{\hsize}{!}{\includegraphics{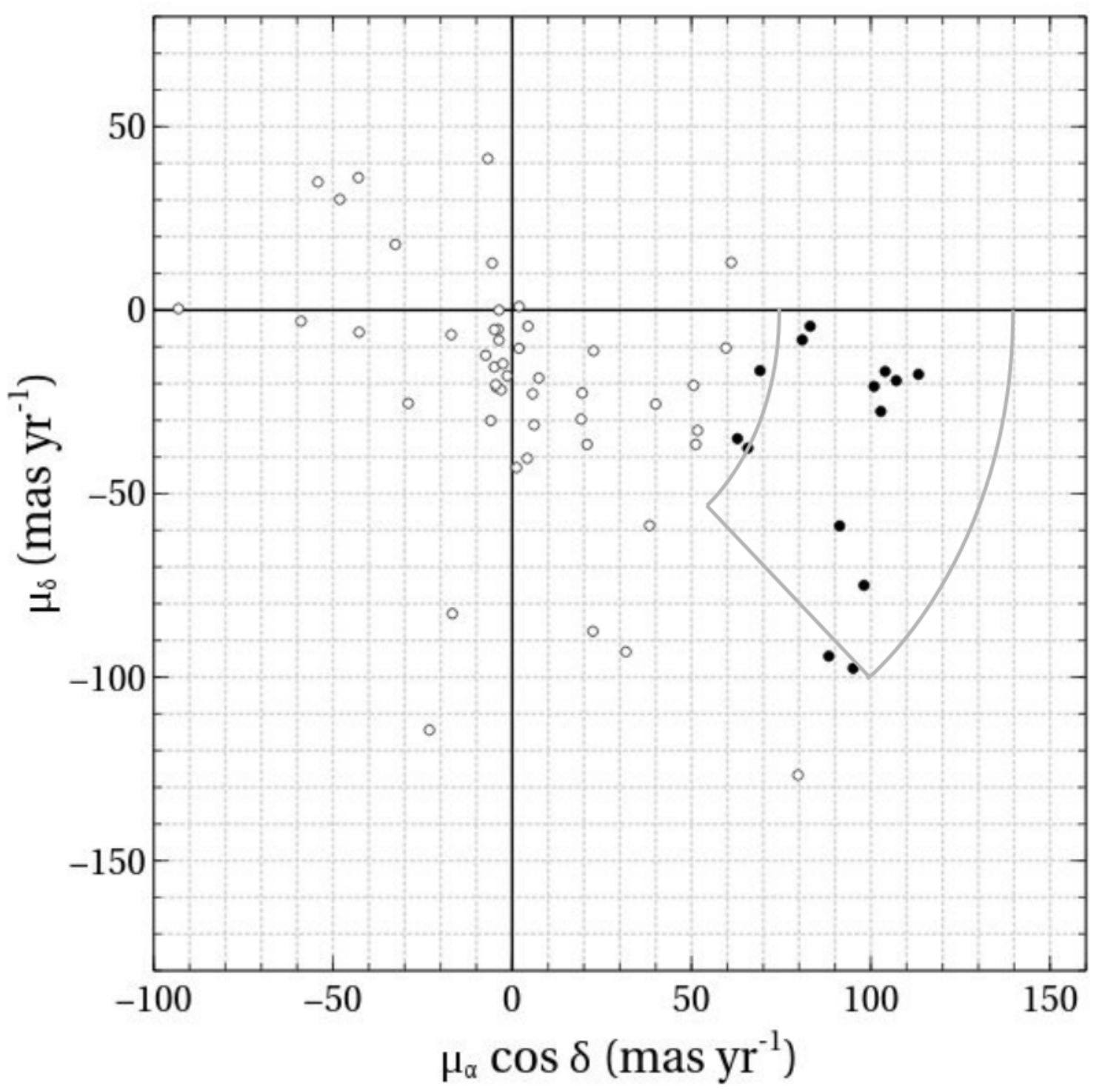}}
%\resizebox{0.5\hsize}{!}{\includegraphics{Hyades_PM_2016_oct.pdf}}
%\includegraphics[scale=0.65]{Hyades_PM_2016.pdf}
%\end{minipage}
\caption{\label{Hyades_pm} PM diagram of photometrically selected objects in the Hyades region as a vector diagram of
$\mu_\alpha\cos\delta$ and $\mu_\delta$. The grey box plotted following \citet{Bou08} and \citet{Bry94} defines the region of PMs expected 
for the cluster. Taking into account rms errors, 14 objects of 66 optically selected candidates (empty circles) were classified as PM members (filled 
circles).}
\end{figure}

Several all-sky surveys exist in the literature which provide PMs for a large number of stars. Cross-identification of our selected candidates in the 
USNO-B1 catalogue show, however, that the given PMs of some objects are questionable because of close stellar binarity that may have affected the 
measurements. The same shortcomings are found in the PPMXL, which  combines USNO-B1.0 and 2MASS astrometry. Therefore, we decided to measure our own 
precise positions for our selected candidates to derive more reliable PM results, especially for the faint objects.

Therefore, we measured the (X-Y) location of our optically selected candidates on the astrometrically calibrated TLS $RI$ images and determined their 
sky coordinates (J2000). To compute the PMs from several epochs, we then cross-referenced our candidates with the following surveys: POSS1 ($R$), 
POSS2 ($R+I$), 2MASS, UKIDSS $K_1$ band, and WISE; we then obtained astrometry for seven independent epochs (TLS $RI-$ averaged) from 1950 to 2010. 
Most objects fainter than $I = 17$ could not be detected on the POSS1-R plates. These objects were measured on the WISE 3.4 and 4.6 $\mu$m bands, 
where they were all detected, and were small enough to provide high-quality astrometry. These objects therefore only have an epoch difference from 
1990 to 2010. We note that all objects could be measured on the downloaded 2MASS images, even those that are not in the 2MASS catalogue, while UKIDSS 
covers the Hyades only in its $K_1$ band.

The equatorial coordinates were extracted from all photometric surveys for the same epoch J2000 and thus a change of object positions should only 
depend on its PM. Finally, the PM of every target was derived from linear fitting the position changes over the epochs spanned by our data. The 
typical errors of the measurements are $\sigma(\mu_\alpha\cos\delta) = 14.1$ mas yr$^{-1}$ and $\sigma(\mu_\delta) = 11.2$ mas yr$^{-1}$ for the PMs 
measured within 1950--2010 and $\sigma(\mu_\alpha\cos\delta) = 28.5$ mas yr$^{-1}$ and $\sigma(\mu_\delta) = 27.2$ mas yr$^{-1}$ for the 1990--2010 
epochs.

One source that can affect our PM results is a visual binary system. \citet{Reid93} found that the Hyades may have a substantial fraction of the 
binaries,  from 25\% to 60\% depending on the model. If we measure the position of an unresolved binary, we believe that we derive the motion of the 
system as a whole,  but if we measure the motion of an object which is actually a companion of a visual binary system, the measured PM can differ from 
that of the system because the components are involved in motion around their centre of mass. We inspected all selected member candidates and if they 
had signs of binarity, we marked them in Tables~\ref{pm_memb} and \ref{pm_nonmemb} as `VB' for probable visual binaries (visual partially resolved 
systems) or as `VB?' for wide stellar pairs with lower probability (resolved systems). One can see that the binary fraction in our selection set is 
not as high as predicted by \citet{Reid93}. This is  probably  due to the  source detection method used;  the method used the shape of stellar 
images to separate single stellar objects from extended sources such as galaxies or artefacts. As a result, the many partially resolved systems whose 
components were not resolved during the detection process seem to have been excluded.

The resulting measured PMs for all 66 objects are shown as a vector diagram of $\mu_\alpha\cos\delta$ and $\mu_\delta$ in Fig.~\ref{Hyades_pm}. The 
objects with PMs satisfying Hyades membership were selected using the same PM box as in \citet{Bou08,Bry94}. Four other objects are located outside of 
this box, but can also be classified as Hyades members within their error bars. In total, 14 objects were selected as Hyades PM members 
(Table~\ref{pm_memb}). Two objects have been selected as possible components of visual binaries TLS-Hy-2 and -8; both have also been classified as 
Hyades members in previous studies as LH 234 \citep{LH88} and RH 230 \citep{Reid92}. However, both objects are mentioned in these studies as single 
stars without  counterparts. Therefore, it is possible that these visual pairs are not physical binaries, but were classified as such due to 
projection effects. The optically selected candidates whose PMs do not satisfy Hyades membership are listed in Table~\ref{pm_nonmemb}. Their PMs are 
listed with individual rms errors and actual epoch range.

\begin{table*}
\small
\caption{\label{pm_memb} Hyades PM member candidates.}
%\begin{tabular}{lccccrrcll}
\begin{tabular}{l@{\hspace{1mm}}c@{\hspace{1mm}}c@{\hspace{1mm}}c@{\hspace{1mm}}c@{\hspace{1mm}}c@{\hspace{1mm}}c@{\hspace{1mm}}c@{\hspace{1mm}}
l@{\hspace{1mm}}l}
\hline\hline
Object    &   RA$_{TLS}$ &   DEC$_{TLS}$ & $I$ & $R-I$ & $\mu_\alpha\cos\delta$ & $\mu_\delta$ & epoch & mass & notes \\
TLS-Hy-.. &   \multicolumn{2}{c}{(J2000)}    & \multicolumn{2}{c}{(mag)} &  \multicolumn{2}{c}{(mas yr$^{-1}$)} &      &   ($M_\odot$)  &       \\
\hline
1         &  04 17 31.3 & 15 23 01 & 14.88 & 1.61 &  95.1$\pm$16.6 & $-97.6\pm26.0$ & 1950.94--2006.91 & 0.12  & \\ 
2         &  04 18 00.5 & 13 35 58 & 16.04 & 2.29 &  69.1$\pm$24.8 & $-16.5\pm$10.0 & 1953.78--2006.91 & 0.09  & LH 234, VB\\
3         &  04 18 51.1 & 13 59 24 & 14.82 & 1.73 &  98.1$\pm$18.6 & $-75.0\pm$9.6  & 1953.78--2006.91 & 0.13  & LH 222n\\
4         &  04 18 57.7 & 16 10 56 & 15.44 & 1.73 &  65.7$\pm$18.5 & $-37.6\pm$17.3 & 1950.94--2006.91 & 0.10  & \\
5         &  04 19 41.8 & 16 45 22 & 17.00 & 2.42 &  88.3$\pm$42.9 & $-94.3\pm$17.8 & 1995.73--2006.91 & 0.08  & LH 214\\
6         &  04 20 50.3 & 13 45 53 & 17.32 & 2.42 &  80.9$\pm$32.8 & $-8.1 \pm$11.4 & 1989.85--2010.66 & 0.08  & LHD 0418+1338\\
7         &  04 22 05.2 & 13 58 47 & 18.63 & 2.46 & 107.1$\pm$37.7 & $-19.2\pm$3.5  & 1995.73--2010.66 & 0.07  & Hy 6 \\
8         &  04 26 19.1 & 17 03 02 & 14.93 & 1.92 & 102.8$\pm$11.5 & $-27.6\pm$9.9  & 1955.94--2007.18 & 0.12  & RH 230, VB\\
9         &  04 27 05.3 & 13 10 33 & 16.44 & 2.03 &  83.1$\pm$12.9 & $ -4.4\pm$8.1  & 1955.95--2006.91 & 0.09  & \\
10        &  04 29 02.9 & 13 37 59 & 15.15 & 2.32 & 100.9$\pm$12.0 & $-20.8\pm$5.4  & 1955.95--2006.91 & 0.11  & LH 91,CFHTHy-16\\
11        &  04 30 04.2 & 16 04 08 & 15.02 & 2.04 &  91.3$\pm$17.2 & $-58.8\pm$14.6 & 1955.95--2006.91 & 0.12  & RH 281, LH 85, CFHTHy-14\\
12        &  04 31 16.4 & 15 00 12 & 14.69 & 1.92 & 104.0$\pm$17.9 & $-16.7\pm$11.2 & 1955.95--2006.91 & 0.13  & LH 68n, CFHTHy-12\\
13        &  04 32 51.2 & 17 30 09 & 17.83 & 2.67 & 113.3$\pm$25.7 & $-17.5\pm$23.0 & 1955.94--2006.91 & 0.08  & LHD 0429+1723\\
14        &  04 33 28.1 & 17 29 32 & 15.56 & 1.83 &  62.8$\pm$15.6 & $-35.0\pm$27.0 & 1955.94--2006.91 & 0.10  & \\
\hline
\end{tabular}

VB = a visual binary (partially resolved system), Cross-identification: CFHTHy = \citet{Bou08}; Hy = \citet{Hog08}; LH = \citet{LH88}; LHD = \citet{Leg94}; RH = 
\citet{Reid92}.
\end{table*}

\section{Results and discussion}
\label{disc}

Our $RI$ survey covered 23.4 deg$^2$ in the core area of the Hyades (Fig.~\ref{area}). We estimated the completeness and the limiting magnitude
of the survey in the $RI$ bands as described in \citet{Cab07}. Our calculation shows that our survey is complete to 21.5 in the $R$ band and 20.5
in the $I$ band. The limiting magnitude of the $RI$ survey is about 1.5--2 mag fainter than the completeness magnitude, i.e. the limiting
magnitude is 23 for the $R$ band and 22.5 for $I$. Since we used the same CCD camera, these values are similar to those of our Coma imaging
survey \citep{Mel12}, but the Hyades ($m-M=3.3$) are closer to the Sun than the Coma open cluster ($m-M=4.7$) and therefore we can
detect objects in the Hyades with an absolute magnitude that is  $\sim$1.5 fainter than in the Coma cluster.

Combining TLS $RI-$photometry with 2MASS IR we have selected 66 photometric candidates (Fig.~\ref{CMDs} and Table~\ref{phot_cand}) using the modern 
BT-Settl theoretical model \citep{Bar15,All16}. The photometric candidates span magnitudes from $I\sim 14.5$ to 18.7, covering the mass range from 
$0.15\,M_\odot$ to $0.07\,M_\odot$ (67 Jupiter masses). The objects with $I=14.5-17.7$ correspond to spectral types from M2 down to L0 based on 
spectro-photometric classification from \citet{Kra07}, whereas the fainter objects ($I>17.7$) should have spectral types later than L0. At the adopted 
distance, the boundary between stellar and substellar objects lies at $I\sim18.1$ in the CMD (dashed horizontal line) assuming a Hyades age of 625 
Myr.

The PM selection allowed us to discriminate the cluster members from field objects with PM lower or higher than that of the cluster. The PM 
selection for our photometric set resulted in 14 PM members, which are listed in Table~\ref{pm_memb} containing our $RI$ photometry of the objects as 
well as their PM values. Coordinates of the objects extracted from the TLS frames are based on the J2000 epoch of the Guide Star Catalog 
v1.2. Only one PM member (TLS-Hy-7) is located well under the substellar borderline ($I>18.1$) and we classify it as a photometric BD.

\subsection{Comparison with previous studies}

Several studies of the Hyades during the last years were focused on improving the census of its lowest mass members. Some of the surveys tried to 
cover as much area as possible around the Hyades core using new all-sky surveys. For example, the \citet{Ros11} and \citet{Gold13} studies are based 
on the PPMXL/Pan-STARRS1 surveys and did all-sky searches for Hyades members in a very wide region covering  $\sim$6500 deg$^2$ around the cluster 
centre (a radius of $\sim$30 pc). Other surveys were focused on the core area of this cluster trying to register substellar members towards the lowest 
masses, e.g. the \citet{Bou08} survey. In our survey we used the same approach and tried to discover the faintest $RI$ objects in the Hyades core. 
Nevertheless, our TLS and \citet{Bou08} surveys do not cover the same area. A comparison of our survey with the spatial coverage of Bouvier's study 
(16 deg$^2$) shows that they overlap over $\sim$60\% ($\sim$13.5 sq. deg), i.e. our imaging survey covers an additional 10 deg$^2$. Thus it is 
complementary to the Bouvier et al. study. 

 The \citet{Ros11} and \citet{Gold13} surveys both used the same technique of kinematic selection (the convergent point method), and \citet{Gold13} 
published a list of 63 additional candidates not included in the \citet{Ros11} list. Since the upper mass limit of our TLS survey is $\sim$0.15 
$M_\odot$, it does not overlap with the lower mass limit of the \citet{Ros11} survey ($0.2\,M_\odot$). To check this we tried to cross-reference both 
our PM candidates and photometric candidates with the \citet{Ros11} list, but, as expected, we found no common objects. The \citet{Gold13} survey has 
a lower mass limit of $0.1\,M_\odot$ so that there could be some common sources. However, since the surface density of low-mass Hyades candidates in 
the survey is quite low in the core region, only 3 of 62 candidates from the \citet{Gold13} list are located within the area covered by our TLS 
fields. Nevertheless, we were able to cross-identify two objects with this survey: TLS-Hy-8 = 66.5793+17.0506 and TLS-Hy-12 (photometric candidate) 
= 67.8182+15.0034. The third object from the list of \citet{Gold13} at $R=13.364$ was too bright for the TLS survey.

\citet{Bou08} selected 22 low-mass probable members based on their photometry and PM. Ten of the objects are brighter than $I=14$ and two BD 
candidates have $I>21.5$. Therefore, the magnitudes of these objects were out of our photometric range. Of the ten objects lying within the TLS 
magnitude range, three objects are located outside of the TLS fields. Among the remaining seven objects there are three candidates in common: 
TLS-Hy-10=CFHTHy-16, TLS-Hy-11=CFHTHy-14, and TLS-Hy-12=CFHTHy-12. Four of the objects were not detected because of special observing conditions: 
CFHTHy-15 and -17 are components of a close double source \citep[RHy 240AB,][]{Reid92}, which was not resolved in our survey photometry, and we 
therefore removed these objects from our lists. The remaining two objects (CFHTHy-18 and -19) are located in the vicinity of very bright stars with 
spikes and strong halos, and therefore the  method used was not able to detect and derive photometry for these objects. A comparison of our PM values 
of the objects in common with those of \citet{Bou08} shows that they are in agreement within the rms errors.  

We also searched for our objects in several earlier surveys covering the Hyades core region: \citet{LH88}, \citet{Reid92}, \citet{Leg94}, and 
\citet{Hog08}. The cross-identification is indicated in the `Notes' column of Table~\ref{pm_memb}. In addition to \citet{Bou08}, TLS-Hy-11 was also 
identified as a Hyades member in \citet{Reid92} (RHy 281) and TLS-Hy-10 was selected as a photometric members in \citet{LH88} (LH 91) and 
\citet{Leg94} (LHD0426+1331); LH 68 (TLS-Hy-12) and 222 (TLS-Hy-3) were marked as non-members (n) in \citet{Luy81} based on their PMs. 
However, our PM values calculated over 50 years of epoch difference agree well with other Hyades member candidates. According to the 
photometric distance obtained from UKIRT $JHK$-photometry \citep{LH88} LH 222 is located outside the Hyades core. However, the UKIRT $JHK$-magnitudes 
of LH 222 are considerably different from 2MASS $JHK$-photometry ($>$1 mag); this may be due to genuine variability of this object or may be a false 
detection. In the case of our $RI$ and the 2MASS photometry, a location of LH 222 on both $I-(R-I)$ and NIR CMDs agrees well with the BT-Settl 
theoretical sequence for the Hyades distance. LH 68 was also selected as a probable member  in \citet{Bou08} based on  the criteria of photometry and PM  
(CFHTHy-12).

We have selected only one object (TLS-Hy-7) lying near the substellar domain ($M\leq0.075\,M_\odot$) and cross-identified it with Hy 6 from 
\citet{Hog08}. Since the photometric error is large at these magnitudes, we are unable to ascertain whether the object is lying above or below this 
boundary. Thus, this candidate may be a BD or instead the lowest mass  stellar member known. \citet{Cas14} observed this object with 
medium-resolution NIR spectroscopy and classified TLS-Hy-7 as a Hyades BD member with spectral type of M8--L2. This spectral classification 
 agrees with our estimation (M9--L0) obtained from the spectro-photometric calibration \citep{Kra07}. 

In total, ten previously selected Hyades members are rediscovered in our TLS survey (marked in the `Notes' column in Table~\ref{pm_memb}). Thus only 
four member candidates from our list are not identified in any of the previous surveys and no new BD candidates have been found.

\subsection{An infrared object of special interest: TLS-Hy-153}

\begin{figure}
\centering
\resizebox{0.49\hsize}{!}{\includegraphics{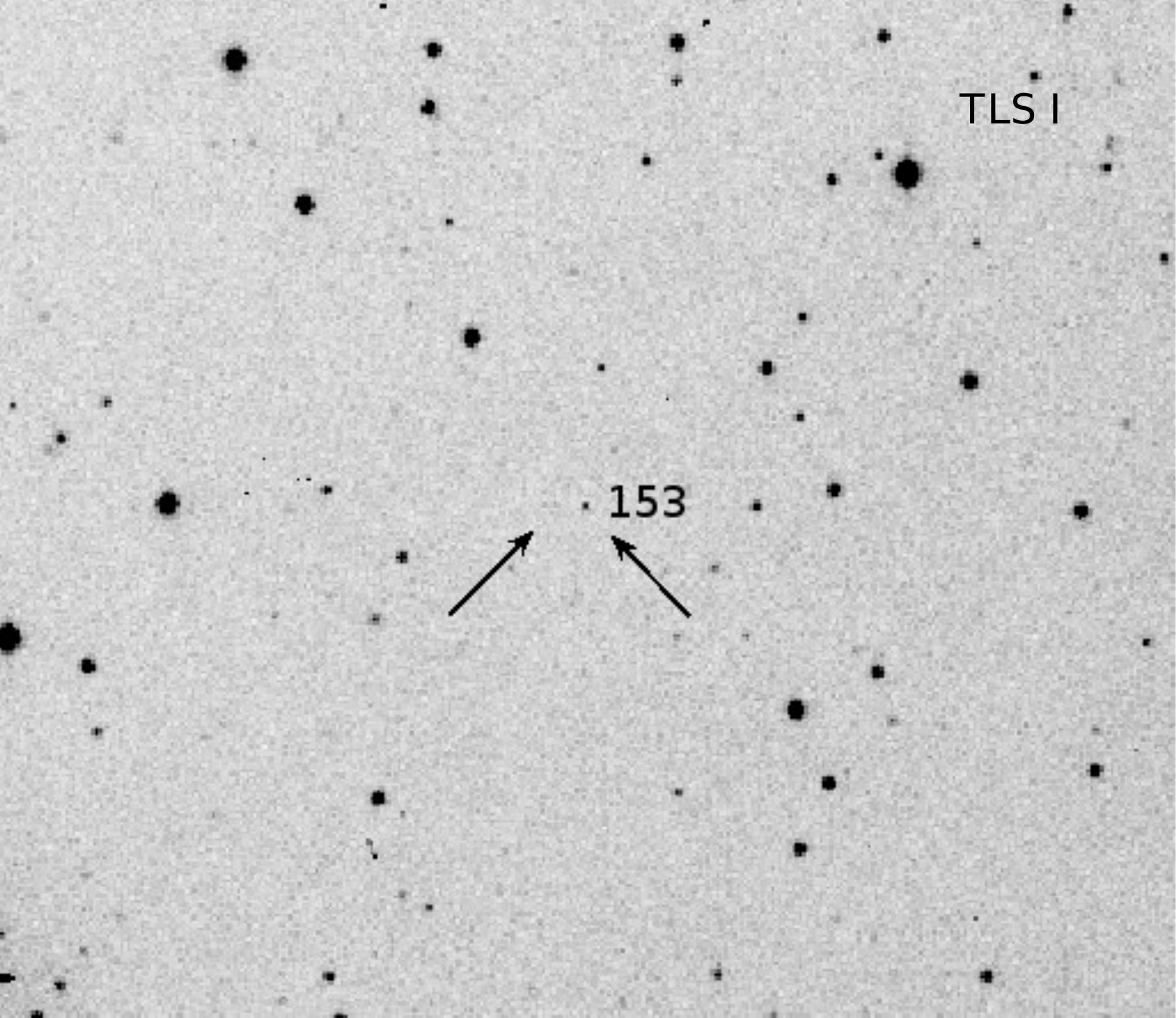}}
\resizebox{0.49\hsize}{!}{\includegraphics{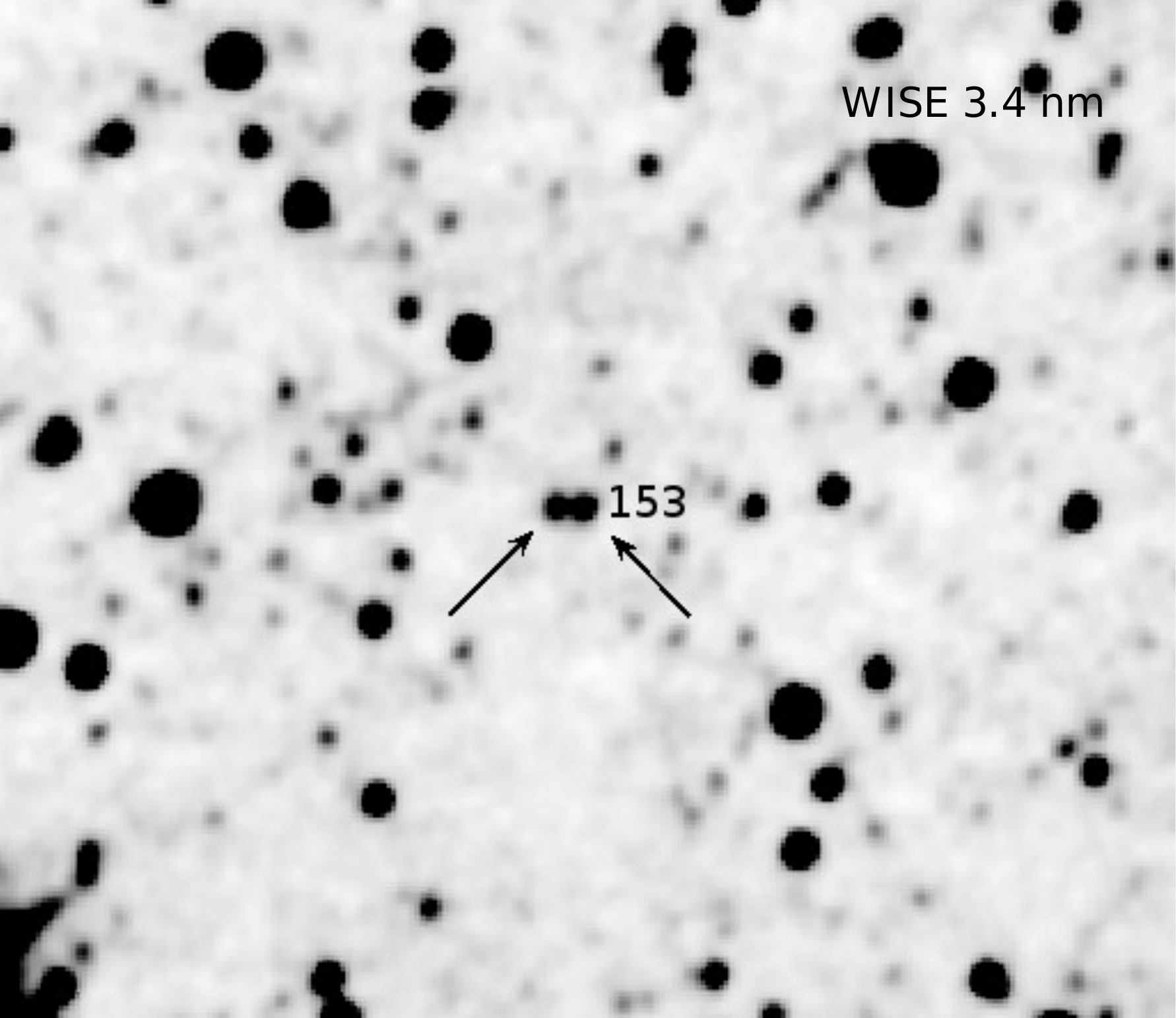}}

\resizebox{0.49\hsize}{!}{\includegraphics{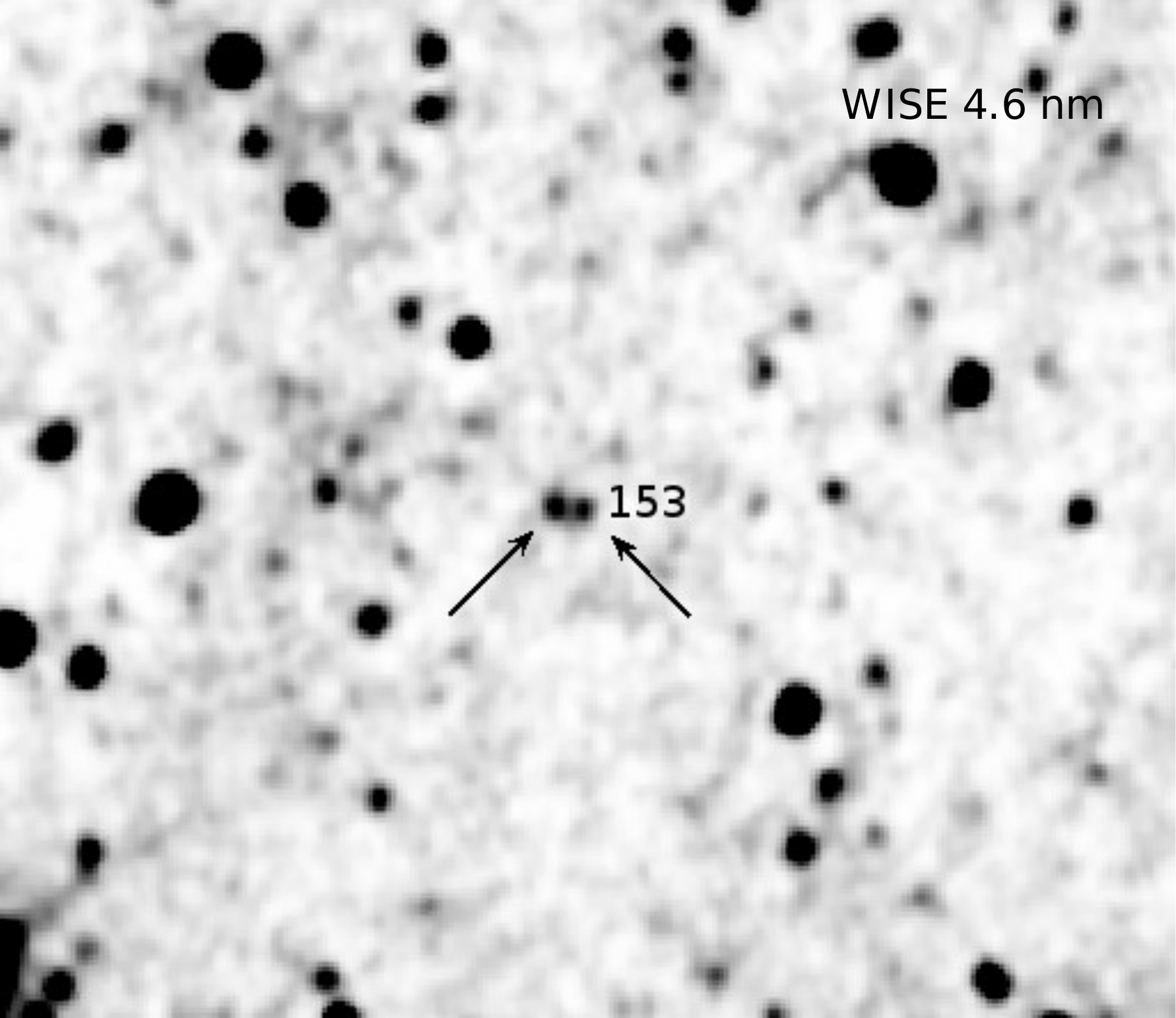}}
\resizebox{0.49\hsize}{!}{\includegraphics{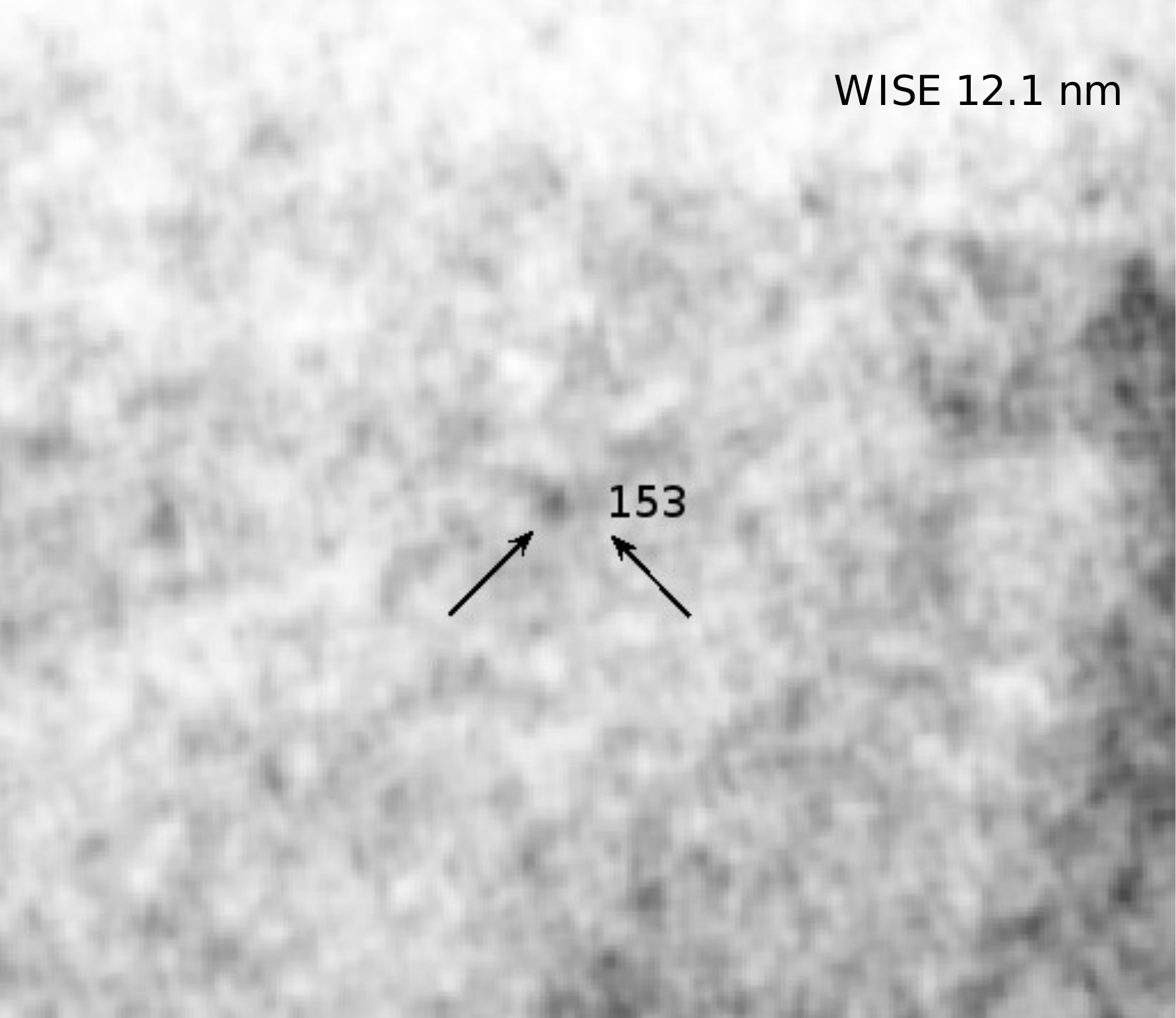}}
%\begin{minipage}[c]{17cm}
%\resizebox{0.7\hsize}{!}{\includegraphics{G7_2097_IR_object.eps}}
%\includegraphics[scale=0.44]{G7_2097_IR_object.eps}
%\end{minipage}
\caption{\label{ir_obj} TLS and WISE images of TLS-Hy-153 region. The IR companion of TLS-Hy-153 (empty arrow) is invisible in the TLS
$I$-band image, but bright in the WISE 3.4 $\mu$m and 4.6 $\mu$m bands. This IR object is also visible at the WISE 12.1 $\mu$m, whereas TLS-Hy-153 is
not visible at the wavelengths.}
\end{figure}

The object TLS-Hy-153 (RA$=4^\mathrm{h}16^\mathrm{m}51^\mathrm{s}$ DEC$=+13\degr\,16\arcmin\,09\arcsec$, $I=19.45$, $R-I=2.72$) was selected as a 
photometric candidate, and it is located in the BD region on the $I-(R-I)$ CMD. Since this object is quite faint, it was not 
cross-identified with a 2MASS source, but only with a UKIDSS object. After transformation from  $K_1$ (UKIDSS) to $K_S$ (2MASS) this object was placed 
on the $I-(I-K)$ CMD, but it was rejected from the list of photometric candidates due to its inconsistent  colour index ($I-K_S = 3.65$). Moreover, the 
derived PM of the object does not agree with that of Hyades. TLS-Hy-153 is hardly visible on the $I$-band image of the TLS survey. At the same time, an 
inspection of WISE images shows that this object is quite bright at 3.4 and 4.6 $\mu$m (Fig.~\ref{ir_obj}) and at UKIDSS $K$ band (2.2 $\mu$m). 
Moreover, the WISE images reveal a bright companion (marked by an arrow) with a similar brightness. At 4.6 $\mu$m this object is brighter than at 3.6 
$\mu$m and it is brighter than TLS-Hy-153 (comparing the peak flux of the central pixels). This IR companion (hereafter TLS-Hy-153-IR) can even be 
seen at 12.1 $\mu$m. This IR object is also detected on the UKIDSS $K$-band image:  J041652.01+131610.0. The angular separation between the two components 
averaged from the high-resolution UKIDSS $K$-map is $\sim$11\arcsec, which is $\sim$500 AU at the distance to the Hyades.

\begin{figure}
\centering
\resizebox{\hsize}{!}{\includegraphics{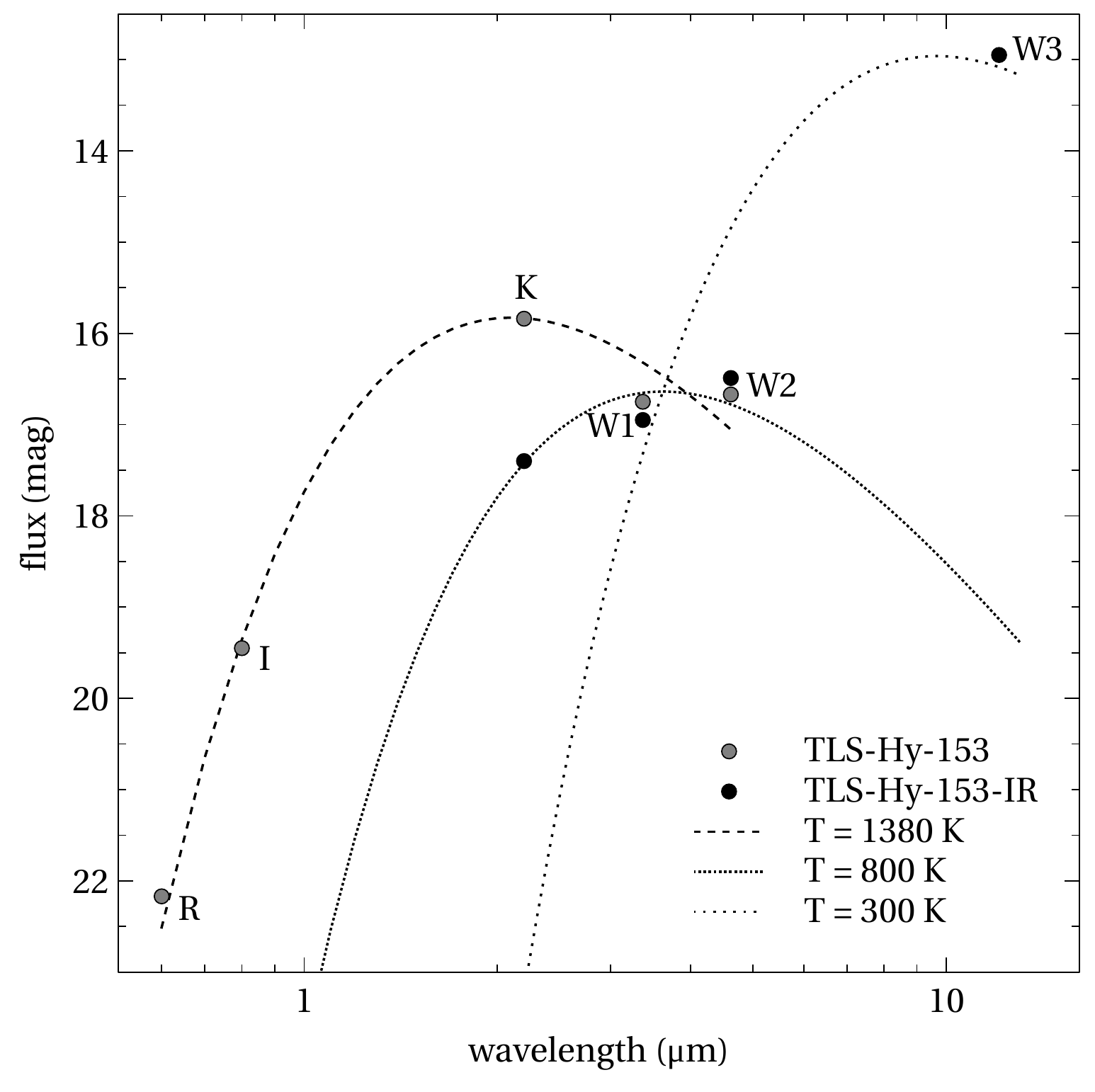}}
\caption{\label{seds} SEDs of TLS-Hy-153 and its IR companion. The TLS-Hy-153 SED is fitted with a single Planck function corresponding to the flux 
distribution of a black body with a temperature of 1380\,K (\textit{dashed line}). The fit is based on the best solution for the blue cut-off and the 
SED slope around the $R,I$ bands at which the stellar object irradiates. The TLS-Hy-153-IR SED is fitted with two Planck functions,  with a black 
body at 800\,K and 300\,K.} \end{figure}

To estimate the temperature of the objects we constructed their SEDs from the  available data: the TLS photometry, the WISE, and the UKIDSS surveys. 
The TLS-Hy-153 SED is based on the five data points measured in the $R,I$ bands; the UKIDSS $K$ band; and WISE W1,W2; whereas photometry data for TLS-Hy-153-IR 
are available from three WISE bands,  W1, W2, and W3 (Fig.~\ref{seds}), and from the UKIDSS $K$ band. The fluxes of the close companion on the 
W1,W2 images  were separated and calculated using PSF-photometry (\textit{IRAF.daophot}), whereas the W3 flux of TLS-Hy-153-IR was estimated using aperture 
photometry; the fluxes were then converted to magnitudes with zero points provided by the WISE survey. Both objects are well resolved on the high-resolution 
UKIDSS $K$-map and therefore we used their photometric magnitudes from the survey.

One can see that TLS-Hy-153 shows a maximum brightness at 2.2 $\mu$m. We estimate the TLS-Hy-153 temperature by fitting its SED with a Planck 
function for a black body. If we assume that the radiation from the  $R,I$ band comes from the stellar object, the best solution that fits the blue cut-off and 
the SED slope around the bands gives a flux distribution at $T=1380$\,K. For TLS-Hy-153-IR, the fit scaled to the IR emission in $K$ and W1--2 corresponds to a 
Planck function with a temperature of $\sim$800\,K, whereas the W3 emission fits better with a black body at 300\,K. This implies that the estimate of the 
TLS-Hy-153-IR temperature using a single Planck function may not be reliable because at these wavelengths different sources with different temperatures can 
contribute to the resultant flux. Indeed, measurements of the diameter of the TLS-Hy-153-IR image in different WISE bands show that the FWHM of its stellar 
profile in the W3 band is larger than in W1 or W2. Therefore, we suggest that TLS-Hy-153-IR represents an extremely faint object (an ultra-cold BD or 
a planet-like object) surrounded by a lower temperature structure such as a dust envelope or a circumstellar disk. 

There are two possibilities to consider for the location of this binary object. The first is that this is a wide binary system located in the Hyades; however,  as 
noted above, because of its faintness the PM of this object could only be obtained over the epoch difference 1997--2010 and does not support its cluster 
membership. On the other hand, the rms of the PM values are quite high. Therefore, we repeated the imaging of this area at the TLS 2m telescope with a longer 
exposure ($2\times$20 min) in the $I$ band in 2015. The astrometry of the brighter companion on these images confirms the previous estimate of its PM direction, 
while the fainter companion is still not visible on these exposures. However, if the system is indeed a wide physical pair, the motion of TLS-Hy-153 around the 
common mass centre may affect the PM determination. Therefore, we cannot exclude the cluster membership of these objects.

A second possibility is that we are looking at a young, more distant system in the Taurus star-forming region (SFR). Since the Hyades are located in 
front of the Taurus SFR, objects in the SFR can be mixed up with genuine Hyades members. We note that the WISE images show a mini-cluster of bright 
IR objects around the young stellar object  (YSO) 2MASS J04220042+1530212 \citep{Reb11}. Many of the objects are not visible on our TLS $R$- and 
$I$-band images, but four objects (TLS-Hy-107, -108, -109, -110) have been selected as photometric candidates in this area; however, they were all 
rejected after the PM selection. Since these objects are projected on the dim area (comparing this region with adjacent ones), all the  IR 
objects might be members of the Taurus SFR. The TLS-Hy-153 system is located 2\fdg5 to the south-east from this mini-cluster and may still belong to 
the periphery of the Taurus SFR. For the Taurus distance of 140 pc, the spatial distance between the objects will be very high: 1540\,AU. On the 
other hand, the star density around TLS-Hy-153 is higher than around this IR cluster  which means that the extinction is not strong in this area. 
However, isolated YSOs located near SFRs in an area without signs of dust clouds with strong extinction are a known phenomenon. Therefore, we neglect 
the interstellar extinction which is probably low in this area. Unfortunately, we do not know a value of the circumstellar extinction for this object 
which might be considerable in the case of YSOs. If we adopt a distance to the SFR as 140 pc \citep{Ken94} and its age as 2 Myr \citep{Ken95}, 
according to the BT-Settl model TLS-Hy-153 is a very low-mass substellar object with $M\sim0.01\,M_\odot$ ($10 M_\mathrm{jup}$) and $T \sim2200$\,K.  
We note, however, that all young BDs that have been found in the Taurus dust clouds so far have been found towards areas with high interstellar 
extinction. Another parameter which we can compare is the PM. The PM of TLS-Hy-153 ($\mu_\alpha\cos\delta=16.1\,\pm\,49.4$ mas yr$^{-1}$, 
$\mu_\delta=-8.1\,\pm\,42.9$ mas yr$^{-1}$) obtained from the short epoch difference coincides within our rms with the PM of other members of the 
Taurus region \citep[e.g. $5.8,-19.5$ mas yr$^{-1}$:][]{Gra13,Duc05}. However, it is very unusual to find such an object so far  from the core of the 
Taurus SFR. If we take the position of the dust cloud where many young stars such as DF Tau and DG Tau are located, 
$\alpha=4^\mathrm{h}$27$^\mathrm{m}$, $\delta$=+25\degr\  50\arcmin, the angular distance between TLS-Hy-153 and this potential birthplace would be 
$\Delta_\alpha\approx3\degr$ and $\Delta_\delta\approx12.5\degr$. To reach the current place within 2 Myr, the difference in velocity between the 
object and the mean Taurus SFR should be about 4 mas yr$^{-1}$ and 22 mas yr$^{-1}$ for $\alpha$ and $\delta$, respectively, which led to 
$\mu_\alpha\cos\delta=1.8$ mas yr$^{-1}$ and  $\mu_\delta=-42$ mas yr$^{-1}$ for TLS-Hy-153. If we take the position of T Tau, which is also 
associated with a dust cloud and close to our target, the difference in the velocity will be less than $\mu_\alpha\cos\delta=3.8$ mas yr$^{-1}$ and  
$\mu_\delta=-31$ mas yr$^{-1}$. Unfortunately, the PM uncertainty for TLS-Hy-153 is quite high, which does not allow us to draw a conclusion on its 
birthplace. Nonetheless, TLS-Hy-153 and the accompanying IR object might represent an interesting wide system:  one of the elements may be a BD and 
the other  a planet-like object (or an ultra-cold BD).

\subsection{Mass function}

The previous studies found that the Hyades are probably a more massive cluster than the similarly aged open cluster in Coma. \citet{Ros11} estimated a cluster 
tidal radius of 9 pc, which is about twice that of Melotte 111 \citep[5--6 pc,][]{Oden98}. Within this tidal radius \citet{Ros11} found 364 stellar systems with 
the total mass of 275 $M_\odot$. \citet{Reid92} estimated the Hyades gravitational binding radius to be as large as $\sim$10.5 pc, comprising a stellar population 
with a total mass of up to 480 $M_\odot$.

The present-day mass function (PDMF) was investigated in detail by \citet{Bou08} based on a large member sample compiled in the Prosser $\&$ Stauffer 
database \citep{Bou08}. This database, combined from many studies, lists more than 500 probable Hyades members and allowed them to build a PDMF 
spanning a range of stellar masses of $0.05-3\,M_\odot$. \citet{Bou08} showed that the Hyades and Pleiades mass functions are similar in shape for 
masses $\geq 1\,M_\odot$ and agree with a Salpeter slope \citep[$\alpha=2.35$,][]{Sal55}. However, for the lower masses the Hyades MF becomes 
shallower than for Pleiades and for a range of $M=0.05-0.2\,M_\odot$ the Hyades MF agrees with a power law index of $\alpha = -1.3$, whereas the 
Pleiades show $\alpha = 0.6$ for the same mass range. This Hyades MF slope has been calculated assuming that the radial distribution of BDs and VLM 
stars is the same and equals $r_C=r_\mathrm{BD}\simeq3$ pc. A larger radius of the BD population ($r_\mathrm{BD}=7$ pc) taken on the assumption of a 
fully relaxed cluster increases the potential amount of BDs, but cannot eliminate the difference between the MFs \citep{Bou08}. This finding confirms 
that the low-mass MF of the Hyades is much more poorly populated than in the Pleiades cluster, which is much younger than the  Hyades.
\begin{figure}
\centering
%\begin{minipage}[c]{17cm}
\resizebox{\hsize}{!}{\includegraphics{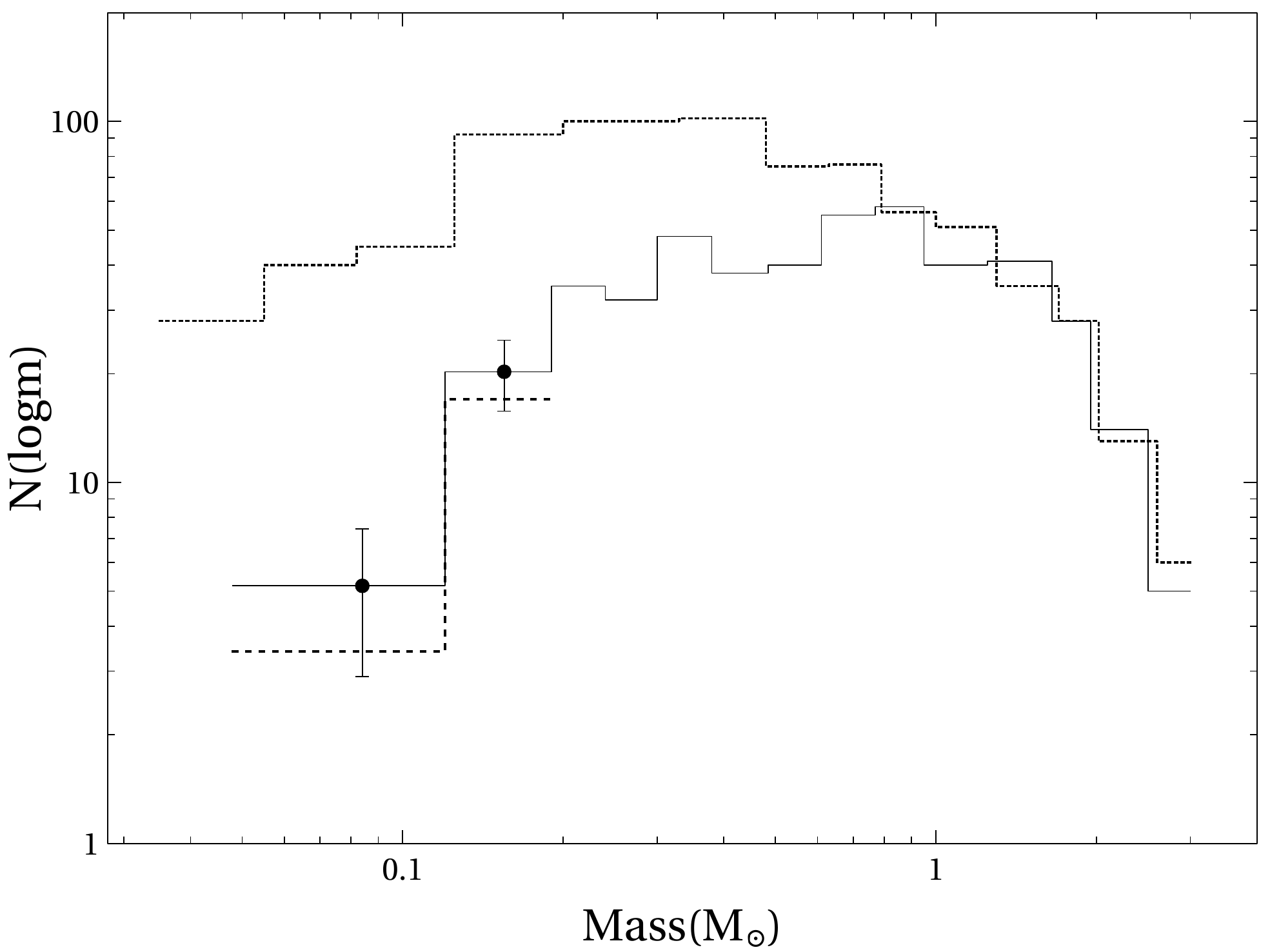}}
%\includegraphics[scale=0.7]{mass_func.eps}
%\end{minipage}
\caption{\label{mass_func} Present-day mass function of the Hyades between 0.05 $M_\odot$ and 3 $M_\odot$. This MF is combined from our new TLS 
survey, the \citet{Bou08} survey and the Prosser \& Stauffer database (solid line). The \textit{dashed} histogram corresponds to the data shown in 
\citet{Bou08}. The data derived from our survey added objects to the Hyades MF in the two lowest mass bins,  0.048--0.12\,$M_\odot$ and 
0.12--0.19\,$M_\odot$. The error bars  take into account the Poissonian error and the photometric uncertainties as well. For comparison, the Pleiades 
MF (\textit{dashed} histogram) adopted from \citet{Bou08} is overplotted. Both the TLS and the Pleiades MFs are normalised to the mass distribution of 
the Prosser \& Stauffer database as described in \citet{Bou08}.} \end{figure}

In order to build an updated PDMF of the Hyades, we combined our new results with the stellar mass statistics presented by \citet{Bou08}. In order to 
calculate the masses for our cluster member selection, we used the mass--magnitude relationships defined by the BT-Settl 625 Myr model \citep{All14}. 
Since the isochrones are available for both $R$- and $I$-magnitudes, we estimated the masses from both bands independently. A comparison of the two 
estimates shows good agreement, within 0.01--0.02\,$M_\odot$. Therefore, we averaged the two values (see  
Table~\ref{pm_memb}).

Figure~\ref{mass_func} represents the resulting new, more complete Hyades PDMF including our data, the compiled published data for 
$M_\star>0.2\,M_\odot$ from the Prosser and Stauffer Open Cluster database, and the data from \citet{Bou08}. Since ten of our objects were 
rediscoveries from previous studies, only four of our new candidates have been added to update the known PDMF. The TLS member set was extrapolated to the 
whole cluster area by a factor, taking into account the ratio of the covered areas in the TLS and Bouvier surveys. The renormalised numbers have been 
added to the final MF. The error bars are based on the Poissonian statistics and also take into account  the mass measurement errors converted from the 
photometric uncertainties. Only one object in our new sample falls within the lowest mass bin (0.048--0.12\,$M_\odot$), whereas three objects have been 
added to the 0.12--0.19\,$M_\odot$ bin. Adding our sample to the Bouvier et al. sample makes the mass spectrum a bit flatter over 
0.05--0.20\,$M_\odot$ and the difference between the Hyades and the Pleiades MFs is still clearly apparent (Fig.~\ref{mass_func}). We should take into 
account that this bin  not only includes BDs, but also the low-mass stellar members. If our selected objects are genuine cluster members, the 
population in the lowest mass bin is more consistent with the core radius of 7 pc for BDs in the Hyades cluster \citep{Bou08}, i.e. $r_C\simeq3$ pc 
for the VLM stars and $r_C\simeq7$ pc for BDs. From the resulting MF over this low-mass range, we find the slope with $\alpha=-1.1\pm0.2,$ which is 
close to the value from \citet{Bou08}.

This result can be explained as a sequence of the continuing dynamical evolution of the Hyades, which are older than the Pleiades. \citet{Age79} showed 
that during the evolution of an open cluster, its members can escape from the cluster and form an extended halo around it. Comparing the shapes of the 
Hyades and Pleiades MFs, \citet{Bou08} estimated that the Hyades must have lost $>$90\% of their initial substellar population ($M<0.08\,M_\odot$).  
However, they concluded that currently $\sim$10--15 BDs could still be located within the Hyades cluster core, whereas initially the cluster 
harboured  up to 200 BDs. Nevertheless, although we used the wide selection criteria which included  the photometric errors and the cluster 
depth and also the potential natural colour dispersion \citep{Reid93}, we did not find any new BD candidates and only four VLM candidates 
were added to the previous list of members. We also note that the area of the TLS survey was 10 deg$^2$ higher than in \citet{Bou08}. Modern numerical 
simulations of the Hyades cluster predict that during the dynamical evolution, `evaporated' BDs and other low-mass members will form elongated 
tails out of the main cluster core \citep{Chum05}. The modelling in \citet{Ern11} shows that the tidal tail of  lost objects can reach a length of 800 
pc after 625 Myr of evolution. Simple calculations show that if the escaping velocity is a few km/sec \citep[for instance, $\sim3$ km/s,][]{Chum05}, 
the former VLM member can recede from the cluster core on several tens of parsecs after 625 Myr of evolution and will be out of our selection 
criteria. Therefore, we cannot exclude that the observed VLM/BD desert might be a sequence of the situation when almost all the VLM/BD members have 
left the cluster core and even its halo.

Therefore, our results support the conclusion that the VLM/BDs deficiency in the Hyades is a consequence of the gradual removal of low-mass cluster 
members due to weak gravitational encounters during the continuous dynamical evolution of the cluster \citep{Bou08}. Moreover, the general 
result from our survey combined with several previous wide surveys shows a lack of any considerable substellar population, which  implies that the 
Hyades core has already lost  most of these objects. The most foreground VLM members probably mixed with foreground dwarfs and migrated so that they 
are no longer  projected on the cluster core. The background evaporated members that are projected onto the core are probably much fainter 
than those that are still located within the cluster volume, and the deeper imaging survey is required to detect such objects.

\section{Conclusion}
We have carried out a wide imaging survey of 23.4 deg$^2$ around the core of the Hyades which partially overlaps with a similar optical survey by 
\citet{Bou08}. Analysis of the TLS $R,I$ photometry, together with 2MASS $JHK_s$ and derived PMs, led to a final list of 14 objects which 
satisfy the membership criteria for the Hyades. We identify four new low-mass member candidates, while a further ten stars from our list can be 
cross-identified with objects discovered in earlier studies. No new photometric substellar objects (BD) were discovered for the 
distance of the Hyades core. We rediscovered only Hy 6 \citep{Hog08} as a PM member and classified it as a photometric substellar 
object candidate (BD) based on the comparison of the observed CMD with theoretical model isochrones. With our four new 
candidates added to the present-day mass function of the Hyades below $0.15\,M_\odot$, the updated mass function is close to that of \citet{Bou08}. 
However, low-resolution spectra of the objects in the red and near-infrared spectral domain are desirable in order to check their ages, which should coincide 
with the cluster age. In the case of a cluster membership, the objects should exhibit signs of relative youth, such as H$\alpha$ in emission 
\citep{Bou08,Mel12}.

\begin{acknowledgements}
J.E. and S.M. acknowledge support from the American Astronomical Society under the 2005 Henri Chr\'etien International Research Grant. This 
publication has  made use of data products from the Two Micron All-Sky Survey, which is a joint project of the University of Massachusetts and the 
Infrared Processing and Analysis Center/California Institute of Technology and the United Kingdom Infrared Deep Sky Survey. This research has made use 
of the SIMBAD database, operated at CDS, Strasbourg, France, and of the IRAF software distributed by NOAO.
\end{acknowledgements}

\bibliographystyle{aa}
\bibliography{paper}

\begin{thebibliography}{61}
\expandafter\ifx\csname natexlab\endcsname\relax\def\natexlab#1{#1}\fi

\bibitem[{{Agekian} \& {Belozerova}(1979)}]{Age79}
{Agekian}, T.~A. \& {Belozerova}, M.~A. 1979, \azh, 56, 9

\bibitem[{{Allard}(2014)}]{All14}
{Allard}, F. 2014, in IAU Symposium, Vol. 299, Exploring the Formation and
  Evolution of Planetary Systems, ed. M.~{Booth}, B.~C. {Matthews}, \& J.~R.
  {Graham}, 271--272

\bibitem[{{Allard}(2016)}]{All16}
{Allard}, F. 2016, in SF2A-2016: Proceedings of the Annual meeting of the
  French Society of Astronomy and Astrophysics, ed. C.~{Reyl{\'e}},
  J.~{Richard}, L.~{Cambr{\'e}sy}, M.~{Deleuil}, E.~{P{\'e}contal},
  L.~{Tresse}, \& I.~{Vauglin}, 223--227

\bibitem[{{Allard} {et~al.}(2012){Allard}, {Homeier}, {Freytag}, \&
  {Sharp}}]{All12}
{Allard}, F., {Homeier}, D., {Freytag}, B., \& {Sharp}, C.~M. 2012, in EAS
  Publications Series, Vol.~57, EAS Publications Series, ed. C.~{Reyl{\'e}},
  C.~{Charbonnel}, \& M.~{Schultheis}, 3--43

\bibitem[{{Baraffe} {et~al.}(1998){Baraffe}, {Chabrier}, {Allard}, \&
  {Hauschildt}}]{Bar98}
{Baraffe}, I., {Chabrier}, G., {Allard}, F., \& {Hauschildt}, P.~H. 1998, \aap,
  337, 403

\bibitem[{{Baraffe} {et~al.}(2003){Baraffe}, {Chabrier}, {Barman}, {Allard}, \&
  {Hauschildt}}]{Bar03}
{Baraffe}, I., {Chabrier}, G., {Barman}, T.~S., {Allard}, F., \& {Hauschildt},
  P.~H. 2003, \aap, 402, 701

\bibitem[{{Baraffe} {et~al.}(2015){Baraffe}, {Homeier}, {Allard}, \&
  {Chabrier}}]{Bar15}
{Baraffe}, I., {Homeier}, D., {Allard}, F., \& {Chabrier}, G. 2015, \aap, 577,
  A42

\bibitem[{{Bertin} \& {Arnouts}(1996)}]{Ber96}
{Bertin}, E. \& {Arnouts}, S. 1996, \aaps, 117, 393

\bibitem[{{Boudreault} {et~al.}(2012){Boudreault}, {Lodieu}, {Deacon}, \&
  {Hambly}}]{Boud12}
{Boudreault}, S., {Lodieu}, N., {Deacon}, N.~R., \& {Hambly}, N.~C. 2012,
  \mnras, 426, 3419

\bibitem[{{Bouvier} {et~al.}(2008){Bouvier}, {Kendall}, {Meeus}, {Testi},
  {Moraux}, {Stauffer}, {James}, {Cuillandre}, {Irwin}, {McCaughrean},
  {Baraffe}, \& {Bertin}}]{Bou08}
{Bouvier}, J., {Kendall}, T., {Meeus}, G., {et~al.} 2008, \aap, 481, 661

\bibitem[{{Bouy} {et~al.}(2015){Bouy}, {Bertin}, {Sarro}, {Barrado}, {Moraux},
  {Bouvier}, {Cuillandre}, {Berihuete}, {Olivares}, \& {Beletsky}}]{Bouy15}
{Bouy}, H., {Bertin}, E., {Sarro}, L.~M., {et~al.} 2015, \aap, 577, A148

\bibitem[{{Bryja} {et~al.}(1994){Bryja}, {Humphreys}, \& {Jones}}]{Bry94}
{Bryja}, C., {Humphreys}, R.~M., \& {Jones}, T.~J. 1994, \aj, 107, 246

\bibitem[{{Caballero} {et~al.}(2007){Caballero}, {B{\'e}jar}, {Rebolo},
  {Eisl{\"o}ffel}, {Zapatero Osorio}, {Mundt}, {Barrado Y Navascu{\'e}s},
  {Bihain}, {Bailer-Jones}, {Forveille}, \& {Mart{\'{\i}}n}}]{Cab07}
{Caballero}, J.~A., {B{\'e}jar}, V.~J.~S., {Rebolo}, R., {et~al.} 2007, \aap,
  470, 903

\bibitem[{{Caffau} {et~al.}(2011){Caffau}, {Ludwig}, {Steffen}, {Freytag}, \&
  {Bonifacio}}]{Caf11}
{Caffau}, E., {Ludwig}, H.-G., {Steffen}, M., {Freytag}, B., \& {Bonifacio}, P.
  2011, \solphys, 268, 255

\bibitem[{{Casewell} {et~al.}(2006){Casewell}, {Jameson}, \& {Dobbie}}]{Cas06}
{Casewell}, S.~L., {Jameson}, R.~F., \& {Dobbie}, P.~D. 2006, \mnras, 365, 447

\bibitem[{{Casewell} {et~al.}(2014){Casewell}, {Littlefair}, {Burleigh}, \&
  {Roy}}]{Cas14}
{Casewell}, S.~L., {Littlefair}, S.~P., {Burleigh}, M.~R., \& {Roy}, M. 2014,
  \mnras, 441, 2644

\bibitem[{{Chabrier} {et~al.}(2000){Chabrier}, {Baraffe}, {Allard}, \&
  {Hauschildt}}]{Cha00}
{Chabrier}, G., {Baraffe}, I., {Allard}, F., \& {Hauschildt}, P. 2000, \apj,
  542, 464

\bibitem[{{Chumak} {et~al.}(2005){Chumak}, {Rastorguev}, \& {Aarseth}}]{Chum05}
{Chumak}, Y.~O., {Rastorguev}, A.~S., \& {Aarseth}, S.~J. 2005, Astronomy
  Letters, 31, 308

\bibitem[{{Copenhagen University} {et~al.}(2006){Copenhagen University},
  {Institute}, {Cambridge}, {Uk}, \& {Real Instituto Y Observatorio de La
  Armada}}]{Cop06}
{Copenhagen University}, O., {Institute}, A.~O., {Cambridge}, {Uk}, \& {Real
  Instituto Y Observatorio de La Armada}, F.~E.~S. 2006, VizieR Online Data
  Catalog, 1304

\bibitem[{de~La~Fuente~Marcos \& de~La~Fuente~Marcos(2000)}]{Fue00}
de~La~Fuente~Marcos, R. \& de~La~Fuente~Marcos, C. 2000, \apss, 271, 127

\bibitem[{{Deluca} \& {Weis}(1981)}]{Del81}
{Deluca}, E.~E. \& {Weis}, E.~W. 1981, \pasp, 93, 32

\bibitem[{{Dobbie} {et~al.}(2002){Dobbie}, {Kenyon}, {Jameson}, {Hodgkin},
  {Hambly}, \& {Hawkins}}]{Dob02}
{Dobbie}, P.~D., {Kenyon}, F., {Jameson}, R.~F., {et~al.} 2002, \mnras, 329,
  543

\bibitem[{{Ducourant} {et~al.}(2005){Ducourant}, {Teixeira}, {P{\'e}ri{\'e}},
  {Lecampion}, {Guibert}, \& {Sartori}}]{Duc05}
{Ducourant}, C., {Teixeira}, R., {P{\'e}ri{\'e}}, J.~P., {et~al.} 2005, \aap,
  438, 769

\bibitem[{{Ernst} {et~al.}(2011){Ernst}, {Just}, {Berczik}, \&
  {Olczak}}]{Ern11}
{Ernst}, A., {Just}, A., {Berczik}, P., \& {Olczak}, C. 2011, \aap, 536, A64

\bibitem[{{Gizis} {et~al.}(1999){Gizis}, {Reid}, \& {Monet}}]{Giz99}
{Gizis}, J.~E., {Reid}, I.~N., \& {Monet}, D.~G. 1999, \aj, 118, 997

\bibitem[{{Goldman} {et~al.}(2013){Goldman}, {R{\"o}ser}, {Schilbach},
  {Magnier}, {Olczak}, {Henning}, {Juri{\'c}}, {Schlafly}, {Chen}, {Platais},
  {Burgett}, {Hodapp}, {Heasley}, {Kudritzki}, {Morgan}, {Price}, {Tonry}, \&
  {Wainscoat}}]{Gold13}
{Goldman}, B., {R{\"o}ser}, S., {Schilbach}, E., {et~al.} 2013, \aap, 559, A43

\bibitem[{{Grankin}(2013)}]{Gra13}
{Grankin}, K.~N. 2013, Astronomy Letters, 39, 336

\bibitem[{{Griffin} {et~al.}(1988){Griffin}, {Griffin}, {Gunn}, \&
  {Zimmerman}}]{Gri88}
{Griffin}, R.~F., {Griffin}, R.~E.~M., {Gunn}, J.~E., \& {Zimmerman}, B.~A.
  1988, \aj, 96, 172

\bibitem[{{Hewett} {et~al.}(2006){Hewett}, {Warren}, {Leggett}, \&
  {Hodgkin}}]{Hew06}
{Hewett}, P.~C., {Warren}, S.~J., {Leggett}, S.~K., \& {Hodgkin}, S.~T. 2006,
  \mnras, 367, 454

\bibitem[{{Hogan} {et~al.}(2008){Hogan}, {Jameson}, {Casewell}, {Osbourne}, \&
  {Hambly}}]{Hog08}
{Hogan}, E., {Jameson}, R.~F., {Casewell}, S.~L., {Osbourne}, S.~L., \&
  {Hambly}, N.~C. 2008, \mnras, 388, 495

\bibitem[{{Jones}(1973)}]{Jon73}
{Jones}, D.~H.~P. 1973, \mnras, 161, 19P

\bibitem[{{Kenyon} {et~al.}(1994){Kenyon}, {Dobrzycka}, \& {Hartmann}}]{Ken94}
{Kenyon}, S.~J., {Dobrzycka}, D., \& {Hartmann}, L. 1994, \aj, 108, 1872

\bibitem[{{Kenyon} \& {Hartmann}(1995)}]{Ken95}
{Kenyon}, S.~J. \& {Hartmann}, L. 1995, \apjs, 101, 117

\bibitem[{{Kraus} \& {Hillenbrand}(2007)}]{Kra07}
{Kraus}, A.~L. \& {Hillenbrand}, L.~A. 2007, \aj, 134, 2340

\bibitem[{{Kroupa}(1995)}]{Kr95}
{Kroupa}, P. 1995, \mnras, 277, 1522

\bibitem[{{Landolt}(1992)}]{Lan92}
{Landolt}, A.~U. 1992, \aj, 104, 340

\bibitem[{{Leggett} {et~al.}(1994){Leggett}, {Harris}, \& {Dahn}}]{Leg94}
{Leggett}, S.~K., {Harris}, H.~C., \& {Dahn}, C.~C. 1994, \aj, 108, 944

\bibitem[{{Leggett} \& {Hawkins}(1988)}]{LH88}
{Leggett}, S.~K. \& {Hawkins}, M.~R.~S. 1988, \mnras, 234, 1065

\bibitem[{{Lodieu} {et~al.}(2014){Lodieu}, {Boudreault}, \&
  {B{\'e}jar}}]{Lod14}
{Lodieu}, N., {Boudreault}, S., \& {B{\'e}jar}, V.~J.~S. 2014, \mnras, 445,
  3908

\bibitem[{{Lodieu} {et~al.}(2012{\natexlab{a}}){Lodieu}, {Deacon}, \&
  {Hambly}}]{Lod12a}
{Lodieu}, N., {Deacon}, N.~R., \& {Hambly}, N.~C. 2012{\natexlab{a}}, \mnras,
  422, 1495

\bibitem[{{Lodieu} {et~al.}(2012{\natexlab{b}}){Lodieu}, {Deacon}, {Hambly}, \&
  {Boudreault}}]{Lod12b}
{Lodieu}, N., {Deacon}, N.~R., {Hambly}, N.~C., \& {Boudreault}, S.
  2012{\natexlab{b}}, \mnras, 426, 3403

\bibitem[{{Luyten} {et~al.}(1981){Luyten}, {Hill}, \& {Morris}}]{Luy81}
{Luyten}, W.~J., {Hill}, G., \& {Morris}, S. 1981, Proper Motion Survey,
  University of Minnesota, 59, 1

\bibitem[{{Melnikov} \& {Eisl{\"o}ffel}(2012)}]{Mel12}
{Melnikov}, S. \& {Eisl{\"o}ffel}, J. 2012, \aap, 544, A111

\bibitem[{{Odenkirchen} {et~al.}(1998){Odenkirchen}, {Soubiran}, \&
  {Colin}}]{Oden98}
{Odenkirchen}, M., {Soubiran}, C., \& {Colin}, J. 1998, \na, 3, 583

\bibitem[{{Pe{\~n}a Ram{\'{\i}}rez} {et~al.}(2012){Pe{\~n}a Ram{\'{\i}}rez},
  {B{\'e}jar}, {Zapatero Osorio}, {Petr-Gotzens}, \& {Mart{\'{\i}}n}}]{Pena12}
{Pe{\~n}a Ram{\'{\i}}rez}, K., {B{\'e}jar}, V.~J.~S., {Zapatero Osorio}, M.~R.,
  {Petr-Gotzens}, M.~G., \& {Mart{\'{\i}}n}, E.~L. 2012, \apj, 754, 30

\bibitem[{{Perryman} {et~al.}(1998){Perryman}, {Brown}, {Lebreton}, {Gomez},
  {Turon}, {Cayrel de Strobel}, {Mermilliod}, {Robichon}, {Kovalevsky}, \&
  {Crifo}}]{Per98}
{Perryman}, M.~A.~C., {Brown}, A.~G.~A., {Lebreton}, Y., {et~al.} 1998, \aap,
  331, 81

\bibitem[{{Rebull} {et~al.}(2011){Rebull}, {Koenig}, {Padgett}, {Terebey},
  {McGehee}, {Hillenbrand}, {Knapp}, {Leisawitz}, {Liu}, {Noriega-Crespo},
  {Ressler}, {Stapelfeldt}, {Fajardo-Acosta}, \& {Mainzer}}]{Reb11}
{Rebull}, L.~M., {Koenig}, X.~P., {Padgett}, D.~L., {et~al.} 2011, \apjs, 196,
  4

\bibitem[{{Reid} \& {Hawley}(1999)}]{Reid99}
{Reid}, I.~N. \& {Hawley}, S.~L. 1999, \aj, 117, 343

\bibitem[{{Reid} \& {Mahoney}(2000)}]{Reid00}
{Reid}, I.~N. \& {Mahoney}, S. 2000, \mnras, 316, 827

\bibitem[{{Reid}(1992)}]{Reid92}
{Reid}, N. 1992, \mnras, 257, 257

\bibitem[{{Reid}(1993)}]{Reid93}
{Reid}, N. 1993, \mnras, 265, 785

\bibitem[{{Robin} {et~al.}(2003){Robin}, {Reyl{\'e}}, {Derri{\`e}re}, \&
  {Picaud}}]{Rob03}
{Robin}, A.~C., {Reyl{\'e}}, C., {Derri{\`e}re}, S., \& {Picaud}, S. 2003,
  \aap, 409, 523

\bibitem[{{R{\"o}ser} {et~al.}(2011){R{\"o}ser}, {Schilbach}, {Piskunov},
  {Kharchenko}, \& {Scholz}}]{Ros11}
{R{\"o}ser}, S., {Schilbach}, E., {Piskunov}, A.~E., {Kharchenko}, N.~V., \&
  {Scholz}, R.-D. 2011, \aap, 531, A92

\bibitem[{{Rossow}(1978)}]{Ros78}
{Rossow}, W.~B. 1978, \icarus, 36, 1

\bibitem[{{Salpeter}(1955)}]{Sal55}
{Salpeter}, E.~E. 1955, \apj, 121, 161

\bibitem[{{Skrutskie} {et~al.}(2006){Skrutskie}, {Cutri}, {Stiening},
  {Weinberg}, {Schneider}, {Carpenter}, {Beichman}, {Capps}, {Chester},
  {Elias}, {Huchra}, {Liebert}, {Lonsdale}, {Monet}, {Price}, {Seitzer},
  {Jarrett}, {Kirkpatrick}, {Gizis}, {Howard}, {Evans}, {Fowler}, {Fullmer},
  {Hurt}, {Light}, {Kopan}, {Marsh}, {McCallon}, {Tam}, {Van Dyk}, \&
  {Wheelock}}]{Skr06}
{Skrutskie}, M.~F., {Cutri}, R.~M., {Stiening}, R., {et~al.} 2006, \aj, 131,
  1163

\bibitem[{{Stauffer} {et~al.}(2007){Stauffer}, {Hartmann}, {Fazio}, {Allen},
  {Patten}, {Lowrance}, {Hurt}, {Rebull}, {Cutri}, {Ramirez}, {Young}, {Rieke},
  {Gorlova}, {Muzerolle}, {Slesnick}, \& {Skrutskie}}]{Sta07}
{Stauffer}, J.~R., {Hartmann}, L.~W., {Fazio}, G.~G., {et~al.} 2007, \apjs,
  172, 663

\bibitem[{{Taylor} \& {Joner}(2002)}]{Tayl02}
{Taylor}, B.~J. \& {Joner}, M.~D. 2002, in Bulletin of the American
  Astronomical Society, Vol.~34, American Astronomical Society Meeting
  Abstracts \#200, 655

\bibitem[{{Terlevich}(1987)}]{Terl87}
{Terlevich}, E. 1987, \mnras, 224, 193

\bibitem[{{Wang} {et~al.}(2014){Wang}, {Chen}, {Lin}, {Pandey}, {Huang},
  {Panwar}, {Lee}, {Tsai}, {Tang}, {Goldman}, {Burgett}, {Chambers}, {Draper},
  {Flewelling}, {Grav}, {Heasley}, {Hodapp}, {Huber}, {Jedicke}, {Kaiser},
  {Kudritzki}, {Luppino}, {Lupton}, {Magnier}, {Metcalfe}, {Monet}, {Morgan},
  {Onaka}, {Price}, {Stubbs}, {Sweeney}, {Tonry}, {Wainscoat}, \&
  {Waters}}]{Wang14}
{Wang}, P.~F., {Chen}, W.~P., {Lin}, C.~C., {et~al.} 2014, \apj, 784, 57

\bibitem[{{Zapatero Osorio} {et~al.}(2014){Zapatero Osorio}, {G{\'a}lvez
  Ortiz}, {Bihain}, {Bailer-Jones}, {Rebolo}, {Henning}, {Boudreault},
  {B{\'e}jar}, {Goldman}, {Mundt}, \& {Caballero}}]{Zap14}
{Zapatero Osorio}, M.~R., {G{\'a}lvez Ortiz}, M.~C., {Bihain}, G., {et~al.}
  2014, \aap, 568, A77

\end{thebibliography}

%\clearpage

\appendix

\onecolumn
\section{Hyades photometric member candidates}
In Table~\ref{phot_cand}, we provide a list of 66 optically selected Hyades member candidates inferred from an $I-(R-I)$ CMD that were
selected for follow-up based on 2MASS $JHK_s$ photometry.

%\begin{longtable}{|c|c|cccc|ccc|ll|}
\begin{longtable}{|@{\hspace{1mm}}c@{\hspace{1mm}} |c@{\hspace{1mm}} |c@{\hspace{1mm}} c@{\hspace{1mm}} c@{\hspace{1mm}} c@{\hspace{1mm}}|c@{\hspace{1mm}} 
c@{\hspace{1mm}} c@{\hspace{1mm}} | l @{\hspace{1mm}}l@{\hspace{1mm}}|}
\caption{$RI$- and 2MASS $JHK_\mathrm{s}$ photometry of Hyades member candidates.\label{phot_cand}}\\
\hline\hline
Object   &   2MASS   & RA$_{TLS}$ & DEC$_{TLS}$ & $I$ & $R-I$ & $J$ & $H$ & $K_\mathrm{s}$ & SpT & SpT \\
\cline{7-9}
TLS-Hy-..&           & \multicolumn{2}{c}{(J2000)} & (mag) & (mag) &\multicolumn{3}{|c|}{2MASS} & $I$ & $JHK_\mathrm{s}$ \\
\hline
\endfirsthead
\caption{continued.}\\
\hline\hline
Object   &   2MASS   & RA$_{TLS}$ & DEC$_{TLS}$ & $I$ & $R-I$ & $J$ & $H$ & $K_\mathrm{s}$ & SpT & SpT \\
TLS-Hy-..&           & \multicolumn{2}{c}{(J2000)} &  (mag) & (mag)   & \multicolumn{3}{|c|}{2MASS} & $I$ & $JHK_\mathrm{s}$ \\
\hline
\endhead
\hline
\endfoot
101 &  04170259+1244198 & 04 17 02.6 & 12 44 20 & 16.15\,$\pm$\,0.01 & 1.95\,$\pm$\,0.02 & 14.02 & 13.45 & 13.19 & M6 & M7  \\
1   &  04173123+1523010 & 04 17 31.3 & 15 23 01 & 14.88\,$\pm$\,0.01 & 1.61\,$\pm$\,0.01 & 13.17 & 12.46 & 12.18 & M5 & M6  \\ 
102 &  04175203+1536240 & 04 17 52.0 & 15 36 24 & 15.65\,$\pm$\,0.01 & 1.90\,$\pm$\,0.01 & 13.53 & 12.91 & 12.64 & M5 & M6  \\
2   &  04180046+1335570 & 04 18 00.5 & 13 35 58 & 16.04\,$\pm$\,0.01 & 2.29\,$\pm$\,0.01 & 14.13 & 13.54 & 13.18 & M6 & M7  \\
103 &  04182141+1651372 & 04 18 21.4 & 16 51 37 & 14.55\,$\pm$\,0.01 & 1.59\,$\pm$\,0.01 & 12.79 & 12.19 & 11.93 & M5 & M5  \\
3   &  04185110+1359240 & 04 18 51.1 & 13 59 24 & 14.82\,$\pm$\,0.01 & 1.73\,$\pm$\,0.01 & 12.87 & 12.30 & 12.00 & M5 & M5  \\
4   &  04185767+1610553 & 04 18 57.7 & 16 10 56 & 15.44\,$\pm$\,0.01 & 1.73\,$\pm$\,0.01 & 13.64 & 13.03 & 12.71 & M5 & M6  \\
104 &  04190236+1305327 & 04 19 02.3 & 13 05 33 & 18.52\,$\pm$\,0.05 & 2.65\,$\pm$\,0.13 & 15.36 & 14.85 & 14.45 & M9 & $>$L0 \\
105 &  04193697+1433329 & 04 19 37.0 & 14 33 33 & 17.06\,$\pm$\,0.01 & 2.36\,$\pm$\,0.02 & 14.36 & 13.67 & 13.26 & M7 & M8  \\ 
5   &  04194169+1645222 & 04 19 41.8 & 16 45 22 & 17.00\,$\pm$\,0.01 & 2.42\,$\pm$\,0.01 & 14.43 & 13.86 & 13.53 & M7 & M8  \\ 
106 &  04195465+1647274 & 04 19 54.7 & 16 47 28 & 15.43\,$\pm$\,0.01 & 1.83\,$\pm$\,0.01 & 13.43 & 12.73 & 12.41 & M5 & M6  \\ 
6   &  04205016+1345531 & 04 20 50.3 & 13 45 53 & 17.32\,$\pm$\,0.02 & 2.42\,$\pm$\,0.04 & 14.27 & 13.56 & 13.06 & M7 & M7  \\ 
107 &  04211753+1530035 & 04 21 17.6 & 15 30 04 & 17.17\,$\pm$\,0.01 & 2.27\,$\pm$\,0.02 & 14.42 & 13.74 & 13.26 & M7 & M8  \\
108 &  04215127+1532560 & 04 21 51.3 & 15 32 57 & 18.42\,$\pm$\,0.05 & 2.20\,$\pm$\,0.09 & 15.46 & 14.04 & 13.46 & M8 & M8  \\
109 &  04215218+1519409 & 04 21 52.2 & 15 19 42 & 16.72\,$\pm$\,0.01 & 2.06\,$\pm$\,0.02 & 14.24 & 13.47 & 13.09 & M6 & M7  \\
7   &  04220512+1358474 & 04 22 05.2 & 13 58 47 & 18.63\,$\pm$\,0.03 & 2.46\,$\pm$\,0.09 & 15.50 & 14.81 & 14.25 & M9 & L0  \\
110 &  04223075+1526310 & 04 22 30.7 & 15 26 32 & 17.23\,$\pm$\,0.02 & 2.32\,$\pm$\,0.04 & 14.36 & 13.55 & 13.08 & M7 & M7  \\
111 &  04223593+1402256 & 04 22 36.0 & 14 02 25 & 15.71\,$\pm$\,0.01 & 1.86\,$\pm$\,0.01 & 13.67 & 12.96 & 12.67 & M5 & M6  \\
112 &  04223914+1657504 & 04 22 39.1 & 16 57 50 & 15.01\,$\pm$\,0.01 & 1.68\,$\pm$\,0.01 & 13.07 & 12.54 & 12.26 & M5 & M6  \\
113 &  04224621+1227080 & 04 22 46.2 & 12 27 09 & 16.08\,$\pm$\,0.02 & 1.95\,$\pm$\,0.03 & 14.04 & 13.48 & 13.13 & M6 & M7  \\
114 &  04225357+1308433 & 04 22 53.6 & 13 08 44 & 14.95\,$\pm$\,0.01 & 1.74\,$\pm$\,0.01 & 12.95 & 12.22 & 11.93 & M5 & M5  \\
115 &  04232421+1559537 & 04 23 24.2 & 15 59 55 & 17.34\,$\pm$\,0.02 & 2.29\,$\pm$\,0.04 & 14.65 & 13.99 & 13.60 & M7 & M9  \\
116 &  04232470+1541451 & 04 23 24.7 & 15 41 43 & 17.28\,$\pm$\,0.02 & 2.41\,$\pm$\,0.04 & 14.67 & 14.12 & 13.79 & M7 & M9  \\
117 &  04232781+1702288 & 04 23 27.8 & 17 02 29 & 15.03\,$\pm$\,0.01 & 1.74\,$\pm$\,0.01 & 13.07 & 12.52 & 12.20 & M5 & M6  \\
118 &  04233547+1552267 & 04 23 35.5 & 15 52 28 & 14.22\,$\pm$\,0.01 & 1.77\,$\pm$\,0.01 & 12.21 & 11.57 & 11.29 & M4 & M5  \\
119 &  04235702+1632458 & 04 23 57.0 & 16 32 46 & 14.16\,$\pm$\,0.01 & 1.69\,$\pm$\,0.01 & 12.27 & 11.72 & 11.44 & M4 & M5  \\
120 &  04240478+1424268 & 04 24 04.8 & 14 24 27 & 15.25\,$\pm$\,0.01 & 1.76\,$\pm$\,0.01 & 13.44 & 12.87 & 12.64 & M5 & M6  \\
121 &  04240618+1439207 & 04 24 06.2 & 14 39 21 & 14.92\,$\pm$\,0.01 & 1.73\,$\pm$\,0.01 & 13.11 & 12.53 & 12.28 & M5 & M6  \\
122 &  04240703+1332409 & 04 24 07.0 & 13 32 41 & 15.80\,$\pm$\,0.01 & 2.06\,$\pm$\,0.01 & 13.38 & 12.82 & 12.51 & M6 & M6  \\
123 &  04243380+1529345 & 04 24 33.8 & 15 29 36 & 15.04\,$\pm$\,0.01 & 2.20\,$\pm$\,0.01 & 12.72 & 12.05 & 11.75 & M5 & M5  \\
124 &  04243861+1604462 & 04 24 38.6 & 16 04 47 & 15.44\,$\pm$\,0.01 & 1.82\,$\pm$\,0.01 & 13.63 & 13.04 & 12.70 & M5 & M6  \\
125 &  04251419+1541079 & 04 25 14.3 & 15 41 08 & 17.70\,$\pm$\,0.02 & 2.69\,$\pm$\,0.04 & 14.71 & 13.94 & 13.49 & M7 & M9  \\
126 &  04252314+1735150 & 04 25 23.1 & 17 35 15 & 17.71\,$\pm$\,0.02 & 2.43\,$\pm$\,0.07 & 14.86 & 14.27 & 13.74 & M7 & L0  \\
127 &  04253933+1723033 & 04 25 39.3 & 17 23 03 & 15.73\,$\pm$\,0.01 & 1.91\,$\pm$\,0.01 & 13.75 & 13.18 & 12.87 & M5 & M5  \\
128 &  04260896+1310379 & 04 26 08.9 & 13 10 38 & 15.59\,$\pm$\,0.01 & 1.83\,$\pm$\,0.01 & 13.40 & 12.72 & 12.38 & M5 & M6  \\
129 &  04261090+1408590 & 04 26 10.9 & 14 08 57 & 15.75\,$\pm$\,0.01 & 2.05\,$\pm$\,0.01 & 13.63 & 13.14 & 12.79 & M6 & M6  \\
8   &  04261903+1703021 & 04 26 19.1 & 17 03 02 & 14.93\,$\pm$\,0.01 & 1.92\,$\pm$\,0.01 & 12.87 & 12.28 & 11.91 & M5 & M5  \\
130 &  04263477+1339146 & 04 26 34.7 & 13 39 14 & 14.98\,$\pm$\,0.01 & 1.74\,$\pm$\,0.01 & 13.18 & 12.58 & 12.28 & M5 & M6  \\
131 &  04264240+1638001 & 04 26 42.4 & 16 38 00 & 15.45\,$\pm$\,0.01 & 1.76\,$\pm$\,0.01 & 13.54 & 12.97 & 12.63 & M5 & M6  \\
132 &  04270289+1558224 & 04 27 02.9 & 15 58 24 & 14.83\,$\pm$\,0.03 & 1.93\,$\pm$\,0.05 & 12.90 & 12.18 & 11.80 & M5 & M5  \\
9   &  04270528+1310323 & 04 27 05.3 & 13 10 33 & 16.44\,$\pm$\,0.01 & 2.03\,$\pm$\,0.01 & 14.32 & 13.71 & 13.36 & M6 & M8  \\
133 &  04273932+1507345 & 04 27 39.2 & 15 07 34 & 15.54\,$\pm$\,0.01 & 2.23\,$\pm$\,0.01 & 13.34 & 12.80 & 12.47 & M5 & M6  \\
134 &  04284199+1533535 & 04 28 42.0 & 15 33 54 & 14.78\,$\pm$\,0.01 & 1.82\,$\pm$\,0.01 & 12.88 & 12.29 & 12.04 & M5 & M5  \\
135 &  04285859+1517386 & 04 28 58.6 & 15 17 38 & 15.07\,$\pm$\,0.01 & 1.95\,$\pm$\,0.01 & 13.06 & 12.38 & 12.08 & M5 & M6  \\
10  &  04290287+1337586 & 04 29 02.9 & 13 37 59 & 15.15\,$\pm$\,0.01 & 2.32\,$\pm$\,0.01 & 12.65 & 11.94 & 11.62 & M5 & M5  \\
11  &  04300417+1604079 & 04 30 04.2 & 16 04 08 & 15.02\,$\pm$\,0.01 & 2.04\,$\pm$\,0.02 & 12.88 & 12.33 & 11.99 & M5 & M5  \\
136 &  04301227+1301086 & 04 30 12.3 & 13 01 09 & 16.48\,$\pm$\,0.02 & 2.07\,$\pm$\,0.03 & 14.14 & 13.54 & 13.16 & M6 & M7 \\
137 &  04302653+1737482 & 04 30 26.5 & 17 37 48 & 15.38\,$\pm$\,0.01 & 1.75\,$\pm$\,0.01 & 13.55 & 12.92 & 12.64 & M5 & M6  \\
138 &  04302914+1347595 & 04 30 29.1 & 13 48 01 & 14.87\,$\pm$\,0.01 & 1.82\,$\pm$\,0.01 & 12.84 & 12.07 & 11.72 & M5 & M5  \\
139 &  04304840+1455210 & 04 30 48.4 & 14 55 21 & 15.10\,$\pm$\,0.01 & 1.68\,$\pm$\,0.01 & 13.31 & 12.64 & 12.33 & M5 & M6  \\
140 &  04305333+1725355 & 04 30 53.3 & 17 25 36 & 14.78\,$\pm$\,0.01 & 1.76\,$\pm$\,0.02 & 13.05 & 12.41 & 12.17 & M5 & M6  \\
141 &  04310937+1706564 & 04 31 09.3 & 17 06 56 & 15.22\,$\pm$\,0.02 & 1.73\,$\pm$\,0.02 & 13.37 & 12.83 & 12.54 & M5 & M6  \\
12  &  04311634+1500122 & 04 31 16.4 & 15 00 12 & 14.69\,$\pm$\,0.01 & 1.92\,$\pm$\,0.01 & 12.63 & 12.07 & 11.71 & M5 & M5  \\
142 &  04312630+1329547 & 04 31 26.3 & 13 29 55 & 15.19\,$\pm$\,0.01 & 1.73\,$\pm$\,0.01 & 13.32 & 12.78 & 12.46 & M5 & M5  \\
143 &  04315853+1300465 & 04 31 58.5 & 13 00 47 & 15.41\,$\pm$\,0.01 & 1.76\,$\pm$\,0.01 & 13.31 & 12.58 & 12.23 & M5 & M6  \\
144 &  04323779+1437237 & 04 32 37.7 & 14 37 25 & 15.49\,$\pm$\,0.01 & 1.80\,$\pm$\,0.01 & 13.75 & 13.21 & 12.81 & M5 & M7  \\
145 &  04323801+1508525 & 04 32 38.0 & 15 08 52 & 16.87\,$\pm$\,0.01 & 2.33\,$\pm$\,0.02 & 13.92 & 13.14 & 12.73 & M7 & M6  \\
13  &  04325119+1730092 & 04 32 51.2 & 17 30 09 & 17.83\,$\pm$\,0.02 & 2.67\,$\pm$\,0.06 & 14.69 & 13.99 & 13.56 & M8 & M9  \\
146 &  04325917+1652587 & 04 32 59.1 & 16 52 58 & 15.11\,$\pm$\,0.01 & 1.74\,$\pm$\,0.01 & 13.23 & 12.54 & 12.24 & M5 & M6  \\
14  &  04332808+1729317 & 04 33 28.1 & 17 29 32 & 15.56\,$\pm$\,0.01 & 1.83\,$\pm$\,0.01 & 13.71 & 13.15 & 12.86 & M5 & M6  \\
147 &  04333831+1712198 & 04 33 38.3 & 17 12 18 & 14.75\,$\pm$\,0.01 & 1.86\,$\pm$\,0.01 & 12.81 & 12.23 & 11.91 & M5 & M5  \\
148 &  04335100+1257033 & 04 33 51.0 & 12 57 05 & 16.11\,$\pm$\,0.01 & 2.11\,$\pm$\,0.02 & 13.86 & 13.29 & 12.95 & M6 & M7  \\
149 &  04335913+1738506 & 04 33 59.0 & 17 38 50 & 15.92\,$\pm$\,0.01 & 1.98\,$\pm$\,0.01 & 13.90 & 13.35 & 12.99 & M6 & M7  \\
150 &  04341755+1711312 & 04 34 17.5 & 17 11 31 & 15.08\,$\pm$\,0.01 & 2.05\,$\pm$\,0.01 & 12.88 & 12.27 & 11.94 & M5 & M5  \\
151 &  04342571+1426147 & 04 34 25.8 & 14 26 15 & 15.88\,$\pm$\,0.01 & 1.99\,$\pm$\,0.01 & 13.84 & 13.31 & 13.01 & M6 & M7  \\
152 &  04361002+1447125 & 04 36 10.0 & 14 47 12 & 17.43\,$\pm$\,0.01 & 2.49\,$\pm$\,0.04 & 14.83 & 14.20 & 13.81 & M7 & L0  \\
\hline
\end{longtable}

\begin{longtable}{|@{\hspace{1mm}}c@{\hspace{1mm}} r@{\hspace{1mm}} r@{\hspace{1mm}} c@{\hspace{1mm}} l@{\hspace{1mm}}| c@{\hspace{1mm}} r@{\hspace{1mm}} 
r@{\hspace{1mm}} c@{\hspace{1mm }} l@{\hspace{1mm}}|}
\caption{\label{pm_nonmemb} Proper motion of Hyades probable non-members.}\\
%\begin{tabular}{lccrlrcccl}
\hline\hline
Object   &   $\mu_\alpha\cos\delta$ & $\mu_\delta$    & epoch & Notes &  Object  &  $\mu_\alpha\cos\delta$ & $\mu_\delta$   & epoch & Notes \\
         &  (mas yr$^{-1}$)      & (mas yr$^{-1}$)    &       &       &          & (mas yr$^{-1}$)         &   (mas yr$^{-1}$) &       &    \\
\hline
\endfirsthead
\caption{PM, continued.}\\
\hline\hline
Object   &   $\mu_\alpha\cos\delta$ & $\mu_\delta$    & epoch & Notes &  Object  &  $\mu_\alpha\cos\delta$ & $\mu_\delta$   & epoch & Notes \\
         &  (mas yr$^{-1}$)         & (mas yr$^{-1}$) &       &       &          & (mas yr$^{-1}$)         &   (mas yr$^{-1}$) &       &    \\
\hline
\endhead
\hline
\endfoot
101 & $ 50.6\,\pm$\,12.4 &$ -20.5\,\pm$\,6.3  & 1953.78--2010.67 &     & 127 &$  22.7\,\pm$\,14.7 &$ -11.1\,\pm$\,7.6  & 1955.94--2007.18 & VB?  \\
102 & $ 38.3\,\pm$\,19.7 &$ -58.7\,\pm$\,23.4 & 1950.94--2006.91 &     & 128 &$  -3.7\,\pm$\,12.0 &$  -0.0\,\pm$\,7.2  & 1955.95--2006.91 &      \\
103 & $-16.7\,\pm$\,20.5 &$ -82.7\,\pm$\,18.5 & 1950.94--2006.91 &     & 129 &$  72.6\,\pm$\,12.7 &$-280.9\,\pm$\,6.1  & 1955.95--2009.69 &      \\
104 & $-58.9\,\pm$\,13.8 &$  -3.0\,\pm$\,33.6 & 1995.73--2010.66 & VB  & 130 &$   1.2\,\pm$\,11.2 &$ -42.9\,\pm$\,5.2  & 1955.95--2009.69 & VB?  \\
105 & $ 61.1\,\pm$\,18.2 &$  13.0\,\pm$\,4.6  & 1953.78--2006.91 &     & 131 &$  -2.6\,\pm$\,11.8 &$ -14.6\,\pm$\,10.4 & 1955.94--2006.91 & VB?  \\
106 & $ 51.2\,\pm$\,20.6 &$ -36.6\,\pm$\,12.3 & 1950.94--2006.91 &     & 132 &$  -3.7\,\pm$\,18.9 &$  -8.2\,\pm$\,16.6 & 1955.94--2006.91 &      \\
107 & $-42.7\,\pm$\,14.8 &$  -6.0\,\pm$\,9.1  & 1955.95--2007.22 &     & 133 &$ -93.1\,\pm$\,14.3 &$   0.4\,\pm$\,6.5  & 1955.95--2007.22 &      \\
108 & $-32.6\,\pm$\,25.1 &$  17.9\,\pm$\,12.9 & 1989.85--2010.67 &     & 134 &$  -3.1\,\pm$\,14.4 &$ -21.7\,\pm$\,13.0 & 1955.95--2007.22 &      \\
109 & $-29.0\,\pm$\,12.0 &$ -25.4\,\pm$\,14.1 & 1955.95--2007.22 &     & 135 &$  79.7\,\pm$\,15.2 &$-126.7\,\pm$\,12.0 & 1955.95--2007.22 &      \\
110 & $-48.1\,\pm$\,29.8 &$  30.2\,\pm$\,17.1 & 1989.85--2010.67 &     & 136 &$  59.6\,\pm$\,16.1 &$ -10.3\,\pm$\,13.9 & 1955.95--2006.91 &      \\
111 & $ 40.0\,\pm$\,13.1 &$ -25.6\,\pm$\,4.5  & 1955.95--2009.69 &     & 137 &$  51.7\,\pm$\,11.0 &$ -32.8\,\pm$\,15.3 & 1955.95--2007.18 & VB?  \\
112 & $ -6.0\,\pm$\,11.2 &$ -30.1\,\pm$\,10.0 & 1955.95--2007.18 &     & 138 &$   6.1\,\pm$\,15.8 &$ -31.3\,\pm$\,9.7  & 1955.95--2006.91 &      \\
113 & $ 19.5\,\pm$\,2.6  &$ -22.6\,\pm$\,3.0  & 1955.95--2010.67 &     & 139 &$   1.9\,\pm$\,12.2 &$ -10.4\,\pm$\,7.5  & 1955.95--2006.91 &      \\
114 & $  1.9\,\pm$\,13.7 &$   1.0\,\pm$\,9.6  & 1955.95--2006.91 & VB  & 140 &$  -7.4\,\pm$\,10.0 &$ -12.3\,\pm$\,15.5 & 1955.95--2007.18 & VB?  \\
115 & $-54.2\,\pm$\,43.6 &$  34.9\,\pm$\,22.5 & 1989.85--2010.67 &     & 141 &$  31.7\,\pm$\,17.5 &$ -93.1\,\pm$\,14.4 & 1955.94--2007.18 & VB?  \\
116 & $ -0.8\,\pm$\,34.3 &$-260.9\,\pm$\,16.6 & 1989.85--2010.67 &     & 142 &$  -4.6\,\pm$\,10.7 &$ -20.3\,\pm$\,6.5  & 1955.95--2006.91 &      \\
117 & $  5.7\,\pm$\,12.4 &$ -22.8\,\pm$\,8.3  & 1955.95--2007.18 &     & 143 &$  -3.9\,\pm$\,9.2  &$  -5.2\,\pm$\,7.5  & 1955.95--2006.91 &      \\
118 & $ -5.0\,\pm$\,14.5 &$ -15.5\,\pm$\,16.6 & 1955.95--2007.22 & VB  & 144 &$   6.1\,\pm$\,13.1 &$  110.1\,\pm$\,9.9 & 1955.95--2010.67 &      \\
119 & $ -1.4\,\pm$\,13.1 &$ -17.9\,\pm$\,11.5 & 1955.95--2006.91 &     & 145 &$  -5.6\,\pm$\,14.1 &$  12.8\,\pm$\,10.3 & 1955.95--2006.91 &      \\
120 & $-42.9\,\pm$\,11.9 &$  36.1\,\pm$\,4.5  & 1955.95--2010.67 &     & 146 &$  -5.0\,\pm$\,15.4 &$  -5.3\,\pm$\,22.4 & 1955.94--2006.91 &      \\
121 & $ -6.8\,\pm$\,15.3 &$  41.3\,\pm$\,4.8  & 1955.95--2007.22 &     & 147 &$ -23.1\,\pm$\,15.3 &$-114.4\,\pm$\,20.2 & 1955.94--2006.91 &      \\
122 & $  4.4\,\pm$\,14.9 &$  -4.4\,\pm$\,6.1  & 1955.95--2009.69 &     & 148 &$  19.2\,\pm$\,21.4 &$ -29.7\,\pm$\,7.0  & 1955.95--2006.91 &      \\
123 & $  4.2\,\pm$\,13.8 &$ -40.4\,\pm$\,16.8 & 1955.94--2007.22 & VB? & 149 &$ -17.0\,\pm$\,4.0  &$  -6.7\,\pm$\,1.0  & 1955.95--2006.91 &      \\
124 & $ 20.9\,\pm$\,12.9 &$ -36.6\,\pm$\,13.1 & 1955.94--2008.84 &     & 150 &$   7.4\,\pm$\,11.1 &$ -18.5\,\pm$\,21.8 & 1955.95--2006.91 &      \\
125 &$  22.5\,\pm$\,9.8  &$ -87.5\,\pm$\,7.6  & 1955.94--2009.68 & VB? & 151 &$ 204.7\,\pm$\,10.1 &$  12.9\,\pm$\,6.3  & 1955.95--2010.67 &      \\
126 &$  -4.4\,\pm$\,7.6  &$ -21.0\,\pm$\,7.4  & 1995.73--2010.67 & VB? & 152 &$-112.3\,\pm$\,19.4 &$ -66.2\,\pm$\,10.7 & 1989.94--2010.68 & VB?  \\
\hline
%\end{tabular}
\end{longtable}
VB = a visual binary (partially resolved system), VB? = a possible visual binary (resolved system)

\end{document}